\documentclass{emulateapj}

\newcommand{\Kepler}{{\it Kepler\ }}
\newcommand{\Porb}{P_{\rm orb}}
\newcommand{\nuorb}{\nu_{\rm orb}}

\newcommand{\pdotplus}{\dot P_+}
\newcommand{\cd}{{\rm\ c\ d^{-1}}}

\slugcomment{To appear in ApJ}

\begin{document}

\title{V344 Lyrae: A Touchstone SU UMa Cataclysmic Variable in the Kepler Field}

\author{Matt A. Wood\altaffilmark{1}, 
Martin D. Still\altaffilmark{2,3}, 
Steve B. Howell\altaffilmark{4,2}, 
John K. Cannizzo\altaffilmark{5,6}
Alan P. Smale\altaffilmark{7}
}

\altaffiltext{1}{Department of Physics and Space Sciences,
        Florida Institute of Technology,
        Melbourne, FL\ \ 32901, USA}
\altaffiltext{2}{NASA Ames Research Center, Moffett Field, CA 94095,
USA}
\altaffiltext{3}{Bay Area Environmental Research Institute, Inc., 560
Third St. West, Sonoma, CA 95476, USA}
\altaffiltext{4}{National Optical Astronomy Observatory, Tucson, AZ
	85719, USA}
\altaffiltext{5}{CRESST and Astroparticle Physics Laboratory
NASA/GSFC, Greenbelt, MD 20771, USA}
\altaffiltext{6}{Department of Physics, University of Maryland,
Baltimore County, 1000 Hilltop Circle, Baltimore, MD 21250, USA}
\altaffiltext{7}{NASA/Goddard Space Flight Center, Greenbelt, MD
20771, USA}

\email{wood@fit.edu}

\begin{abstract}
We report on the analysis of the \Kepler short-cadence (SC) light curve of
V344 Lyr obtained during 2009 June 20 through
2010 Mar 19 (Q2--Q4).  The system is an SU UMa star showing dwarf nova 
outbursts and superoutbursts, and promises to be a touchstone
for CV studies for the foreseeable future.  The system displays both
positive and negative superhumps with periods of 2.20 and 2.06-hr,
respectively, and we identify an orbital period of 2.11-hr.  
The
positive superhumps have a maximum amplitude of $\sim$0.25-mag, the
negative superhumps a maximum amplitude of $\sim$0.8 mag,
and the orbital period at quiescence has an amplitude of $\sim$0.025 mag.
The quality of the \Kepler data is such that we can test
vigorously the models for accretion disk dynamics that have been 
emerging in the past several years.
The SC data for V344 Lyr are consistent with the model
that two physical sources yield positive superhumps: early in the
superoutburst, the superhump signal is generated by viscous
dissipation within the periodically flexing disk, but late in the
superoutburst, the signal is generated as the accretion stream bright
spot sweeps around the rim of the non-axisymmetric disk.  The disk
superhumps are roughly anti-phased with the stream/late superhumps. 
The V344 Lyr data also reveal negative superhumps arising
from accretion onto a tilted disk precessing in the retrograde
direction, and suggest that negative superhumps may appear
during the decline of DN outbursts.  The period of negative
superhumps has a positive $\dot P$ in between outbursts.  
\end{abstract}

\keywords{novae, cataclysmic variables; dwarf novae; stars: individual
(V344 Lyr); white dwarfs; hydrodynamics}

\section{Introduction}

Cataclysmic variable (CV) binary systems typically consist of low-mass
main sequence stars that transfer mass though the L1 inner Lagrange
point and onto a white dwarf primary via an accretion disk.  Within
the disk, viscosity acts to transport angular momentum outward in
radius, allowing mass to move inward and accrete onto the primary
white dwarf \citep[e.g.][]{warner95,fkr02,hellier01}.  In the case of
steady-state accretion the disk is the brightest component of the
system, with a disk luminosity $L_{\rm disk} \sim GM_1 \dot M_1/R_1$,
where $\dot M_1$ is the mass accretion rate onto a white dwarf of mass
$M_1$ and radius $R_1$. 

While members of the novalike (NL) CV subclass display a nearly
constant mean system luminosity, members of the dwarf nova (DN)
subclass display quasi-periodic outbursts of a few magnitudes thought
to arise from a thermal instability in the disk.  Specifically, 
models suggest a heating wave rapidly transitions the disk to a
hot, high-viscosity state which significantly enhances $\dot M_1$ for a
few days.  Furthermore, within the DN subclass there are the SU UMa
systems that in addition to normal DN outbursts display superoutbursts
which are up to a magnitude brighter and last a few times longer than
the DN outbursts.  The SU UMa stars are characterized by the
appearance at superoutburst of periodic large-amplitude photometric
signals (termed {\it positive superhumps}) with periods a few percent
longer than the system orbital periods.  So-called {\it negative}
superhumps (with periods a few percent shorter than $\Porb$) are also
observed in some SU UMa systems.  

The oscillation modes (i.e., eigenfrequencies) of any physical object
are a direct function of the structure of that object, and thus an
intensive study of SU UMa superhumps that can make use of both a
nearly-ideal time-series data set as well as detailed
three-dimensional high-resolution numerical models has the potential
to eventually unlock many of the long-standing puzzles in accretion
disk physics.  For example, a fundamental question in astrophysical
hydrodynamics is the nature of viscosity in differentially rotating
plasma disks.  It is typically thought to result  from the
magnetorotational instability (MRI) proposed by \citet{bh98,balbus03},
but the observations to-date have been insufficient to test the model.

\subsection{V344 Lyrae}

The \Kepler field of view includes 12 CVs in the \Kepler Input Catalog
(KIC) that have published results at the time of this writing.  Ten
(10) of these systems are listed in Table 1 of \citet[hereafter Paper
I]{still10}.   Two additional systems have been announced since that
publication, the dwarf nova system BOKS-45906 (KIC 9778689)
\citep{feldmeier11}, and the AM CVn star SDSS J190817.07$+$394036.4
(KIC 4547333) \citep{fontaine11}.

The star V344 Lyr (KIC 7659570) is a SU UMa star that lies in the
\Kepler field.  \citet{kato93} observed the star during a
superoutburst ($V\sim14$), and reported the detection of superhumps
with a period $P = 2.1948\pm 0.0005$ hr.  In a later study
\citet{kato02} reported that the DN outbursts have a recurrence
timescale of $16\pm3$ d, and that the superoutbursts have a recurrence
timescale of $\sim$110 d.  \citet{ak08} estimated a distance of 619 pc
for the star using a period-luminosity relationship.  

In Paper I we reported preliminary findings for V344 Lyr based on the
second-quarter (Q2) \Kepler observations, during which \Kepler
observed the star with a $\sim$1-min cadence, obtaining over 123,000
photometric measurements.  In that paper we reported on a periodic
signal at quiescence that was either the orbital or negative superhump
period, and the fact that the positive superhump signal persisted into
quiescence and through the following dwarf nova outburst. 

In \citet[hereafter Paper II]{cannizzo10} we presented time-dependent
modeling based on the accretion disk limit cycle model for the 270 d
(Q2--Q4) light curve of V344 Lyr.  We reported that the main decay of
the superoutbursts is nearly perfectly exponential, decaying at a rate
of $\sim$12 d mag$^{-1}$, and that the normal outbursts display a
decay rate that is faster-than-exponential. In addition, we noted that
the two superoutbursts are initiated by a normal outburst. Using the
standard accretion disk limit cycle model, we were able to reproduce
the main features of the outburst light curve of V344 Lyr.  We
significantly expand on this in \citet{cannizzo11} where we present
the 1-year outburst properties of both V344 Lyr and V1504 Cyg
(Cannizzo et al.\ 2011). 

In this work, we report in detail on the results obtained by studying
the \Kepler Q2--Q4 data, which comprise without question the
single-best data set obtained to-date from a cataclysmic variable
star. The data set reveals signals from the orbital period as well as
from positive and negative superhumps.  

\section{Review of Superhumps and Examples}

Before digging into the data, we briefly review the physical
processes that lead to the photometric modulations termed superhumps.  

\subsection{Positive superhumps and the two-source model}

The accretion disk of a typical dwarf nova CV that is in quiescence
has a low disk viscosity and so inefficient exchange of angular
momentum.  As a result, the mass transfer rate $\dot M_{\rm L1}$
through the inner Lagrange point L1 is higher than the mass transfer
rate $\dot M_1$ onto the primary.  Thus, mass accumulates in the disk
until a critical surface density is reached at some annulus, and the
fluid in that annulus transitions to a high-viscosity state
\citep{cannizzo98,cannizzo10}.  This high-viscosity state propagates
inward and/or outward in radius until the entire disk is in a
high-viscosity state characterized by very efficient angular momentum
and mass transport -- the standard DN outburst \citep[see, e.g.,][for
reviews]{cannizzo93,lasota01}.  In this state, $\dot M_1 > \dot M_{\rm
L1}$ and the disk drains mass onto the primary white dwarf.

During each DN outburst, however, the angular momentum transport acts
to expand the outer disk radius slightly, and after a few to several
of these, an otherwise normal DN outburst can expand the outer radius
of the disk to the inner Lindblad resonance (near the 3:1 corotation
resonance).  This can only occur for systems with mass ratios
$q=M_2/M_1 \lesssim 0.35$ \citep{wts09}.

Once sufficient mass is present at the resonance radius, the common
superhump oscillation mode can be driven to amplitudes that yield
photometric oscillations.  The superhump oscillation has a period
$P_+$ which is a few percent longer than the orbital period, where the
{\it fractional period excess} $\epsilon_+$ is defined as
\begin{equation}
\epsilon_+\equiv {P_+-\Porb\over\Porb}.
\label{eq: eps+}
\end{equation}
These are the are the so-called {\it common} or {\it positive}
superhumps, where the latter term reflects the sign of the period excess
$\epsilon_+$.  In addition to the SU UMa stars, positive superhumps
have also been observed in novalike CVs
\citep{pattersonea93b,retterea97,skillmanea97,patterson05,kim09}, the interacting binary
white dwarf AM CVn stars
\citep{pattersonea93a,warner95amcvn,nelemans05,roelofs07,fontaine11}, and in low-mass X-ray
binaries \citep{charlesea91,mho92,oc96,retterea02,hynesea06}.  

\begin{figure*}
\plotone{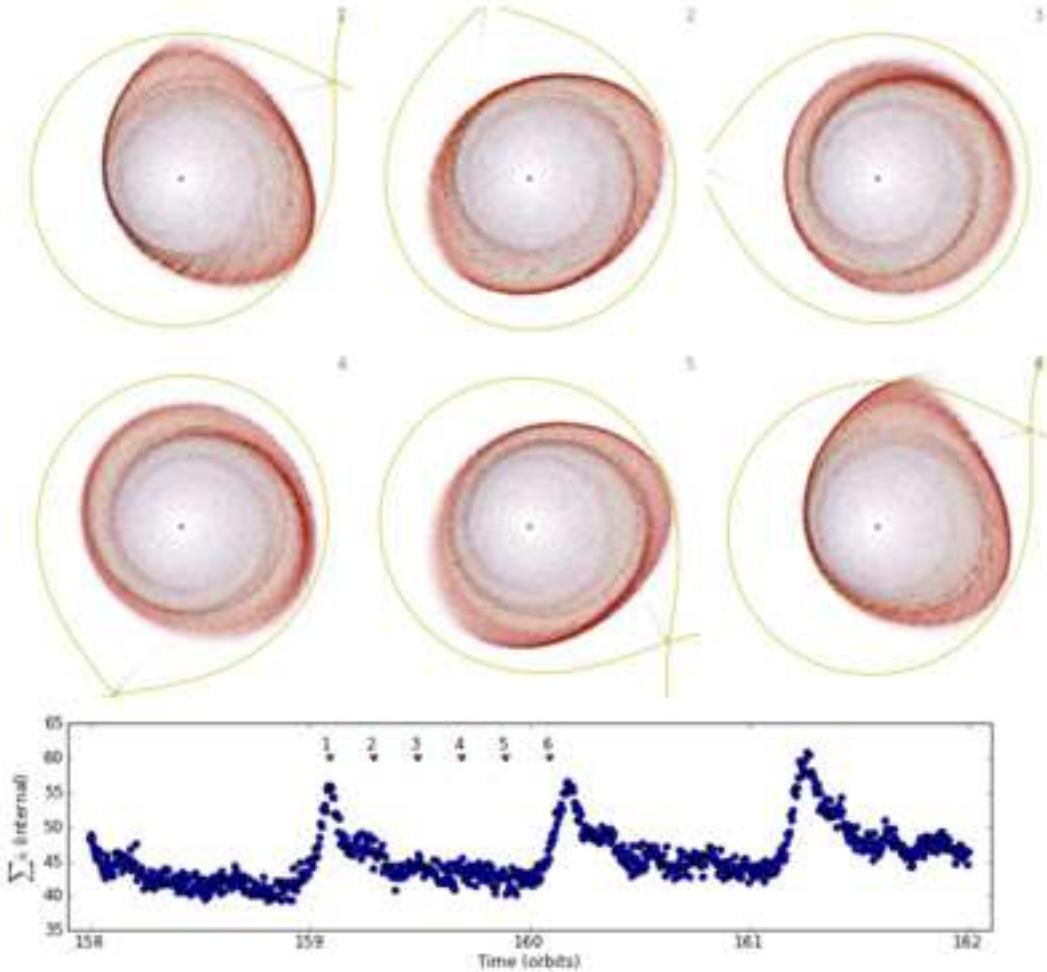}
\caption{Snapshots from one orbit of a $q=0.25$ SPH accretion disk simulation shortly after superhump onset.
The 100,000 simulation particles are color-coded by their internal
energy change (``luminosity'') over the previous simulation time step.
The simulation light curve shown is calculated as the sum of the
luminosities of all the particles.  Note that superhump maximum
corresponds to frames 1 and 6, where the disk opposite the secondary
is radially compressed and hence a region of strongly convergent flows.  Note also that the accretion stream impact region is deeper in the potential well in frame 3 than in frames 1 and 6 -- this yields the ``late'' superhump signal.}
\label{fig: sph+}
\end{figure*}

Figure \ref{fig: sph+} shows snapshots from one full orbit of a
smoothed particle hydrodynamics (SPH) simulation ($q=0.25$, 100,000
particles) as well as the associated simulation light curve
\citep[see][]{sw98,wb07,wts09}.   The disk particles are color-coded
by the change in internal energy over the previous timestep, and the
Roche lobes and positions of $M_1$ are also shown.  Panels 1 and 6 of
Figure \ref{fig: sph+} shows the geometry of the disk at superhump
maximum.  Note that here the superhump light source is viscous
dissipation resulting from the compression of the disk opposite the
secondary star.  The local density and shear in this region are both
high, leading to enhanced viscous dissipation in the strongly
convergent flows.  The orbit sampled in the Figure is characteristic
of early superhumps where the disk oscillation mode is saturated, and
the resulting amplitude significantly higher ($\sim$0.15 mag) than the
models produce when dynamical equilibrium ($\sim$0.03 mag) due to the
lower mean energy production in the models at superhump onset.  

As a further detail, we note that whereas the 2 spiral dissipation
waves are stationary in the co-rotating frame before the onset of the
superhump oscillation, once the oscillation begins, the spiral arms
advance in the prograde direction by $\sim$180$^\circ$ in the
co-rotating frame during each superhump cycle.  This prograde
advancement can be seen by careful inspection of the panels in Figure
\ref{fig: sph+}.   Indeed, this motion of the spiral dissipation waves
is central to the superhump oscillation -- a spiral arm is ``cast''
outward as it rotates through the tidal field of the secondary, and
then brightens shortly afterward as it compresses back into the disk
in a converging flow \citep{smith07,wts09}.  

While viscous dissipation within the periodically-flexing disk
provides the dominant source of the superhump modulation, the
accretion stream bright spot also provides a periodic photometric
signal when sweeping around the rim of a non-axisymmetric disk
\citep{vogt82,osaki85,whitehurst88,kunze04}.   The
bright spot will be most luminous when it impacts most deeply in the
potential well of the primary (e.g., panel 3 of Figure~\ref{fig:
sph+}, and fainter when it impacts the rim further from the white
dwarf primary (Panels 1 and 6).  This signal is swamped by the
superhumps generated by the flexing disk early in the superoutburst,
but dominates once the disk is significantly drained of matter and
returns to low state.  The disk will continue to oscillate although
the driving is much diminished, and thus the stream mechanism will
continue to yield a periodic photometric signal of decreasing
amplitude until the oscillations cease completely.   

This photometric signal is what is termed {\it late superhumps} in the
literature \citep[e.g.,][]{hessman92,patterson00,patterson02,
templeton06,sterken07,kato09,kato10}.  \citet{rolfe01} presented a
detailed study of the deeply eclipsing dwarf nova IY UMa observed
during the late superhump phase where they found exactly this
behavior.  They used the shadow method \citet{wood86} to determine the
radial location of the bright spot (disk edge) in 22 eclipses observed
using time-series photometry.   They found that the disk was
elliptical and precessing slowly at the beat frequency of the orbital
and superhump frequencies, and that the brightness of the stream-disk
impact region varied as the square of the relative velocity of the
stream and disk material \citep[see also][]{smak10}.  Put another way,
the bright spot was brighter when it was located on the periastron
quadrant of the elliptical disk, and fainter on the apastron quadrant.

Thus, two distinct physical mechanisms give rise to positive
superhumps: viscous dissipation in the flexing disk, driven by the
resonance with the tidal field of the secondary, and the time-variable
viscous dissipation of the bright spot as it sweeps around the rim of
a non-axisymmetric disk\footnote{For completeness, we note that
recently Smak (2009, 2011) has proposed that the standard model,
described above, does not explain the physical source of observed
superhump oscillations.  Instead, he suggests that irradiation on the
face of the secondary is modulated, which yields a modulated mass
transfer rate $\dot M_{\rm L1}$, which in turn results in modulated
dissipation of the kinetic energy of the stream.}.  For the
remainder of this paper we refer to this as the {\it two-source model
of positive superhumps} \citep[see also][]{kunze02,kunze04}.  These
two signals are approximately antiphased, and in systems where both
operate at roughly equal amplitude, the Fourier transform of the light
curve can show a larger amplitude for the second harmonic (first
overtone) than for the fundamental (first harmonic).   

As an example of this double humped light curve, in Figure~\ref{fig:
en400420} we show 20 orbits of the $q=0.25$ simulation discussed above
(Figure~\ref{fig: sph+}) starting at orbit 400, by which time the
system had settled into a state of dynamical equilibrium.  The inset
in this Figure shows the the average superhump pulse shape obtained
from orbits 400-500 of the simulation, where we have set phase zero to
primary minimum.  Note that here the average pulse shape is complex
but approximately double-peaked.  The Fourier transform displays
maximum power at twice the fundamental frequency.  When we examine the
disk profiles, we find that the dominant peak arises from the disk
superhump described above, but the secondary peak roughly half a cycle
later results from the impact of the bright spot deeper in the
potential well of the primary (see panel 4 of Figure 1).  The
substructure of this secondary maximum results from the interaction of
the accretion stream with the spiral arm structures that advance
progradely in the co-rotating frame.  Panel 3
of Figure~\ref{fig: sph+} is representative of the disk structure at
the time of the the small dip in brightness observed at superhump
phase 0.55.  The dip is explained by the fact that the accretion
stream bright spot at this phase is located in the low-density
inter-arm region, and therefore that the accretion stream can
dissipate its energy over a longer distance.  In addition the
oscillating disk geometry results in this region having a larger
radius, and lower velocity contrast near this phase.  \citet{howell96}
discuss the observation and phase evolution of the two secondary humps
in the SU UMa system TV Corvi.

\begin{figure}
\epsscale{1.0}
\plotone{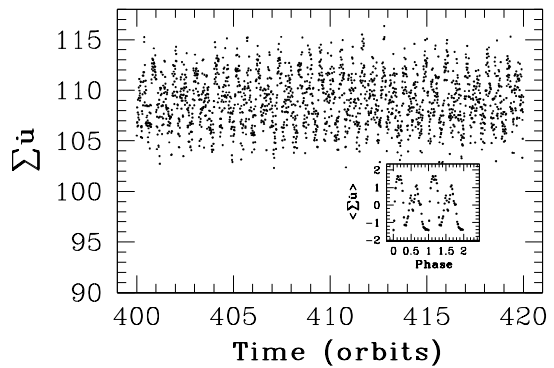}
\caption{Twenty orbits of the simulation light curve once the dynamical
equilibrium state has been reached in the $q=0.25$ simulation
discussed above.  The light curve is double humped.  The inset shows
the average light curve calculated from orbits 400 to 500, 
with the phase set to zero
for primary minimum.  Primary maximum again corresponds to modulation
in the disk as discussed.  Secondary maximum is caused by the
accretion stream impact spot hitting minimum radius of the
non-axisymmetric disk.  The substructure of the secondary maximum is
discussed in the text.  } 
\label{fig: en400420} 
\end{figure}

The 3 AM CVn (helium CV) systems that are in permanent high state --
AM CVn \citep{skillman99}, HP Lib \citep{patterson02} and the system
SDSS J190817.07+394036.4 (KIC 004547333) announced recently by
\citet{fontaine11} -- all display average pulse shapes that are
strongly double humped.  AM CVn itself is frequently observed to show
no power in the Fourier transform at the fundamental superhump
oscillation frequency \citep{smak67,ffw72,patterson92,skillman99}.  AM
CVn systems are known to be helium mass transfer systems with orbital
periods ranging between 5 min and $\sim$1 hr \citep[see reviews
by][]{warner95amcvn,solheim10}.  

In contrast, the hydrogen-rich old-novae and novalike CVs that show
permanent superhumps display mean pulse shapes that are nearly always
similar to the saturation phase light curves as shown in
Figure~\ref{fig: sph+}, and there is no example we know of where a
permanent superhump system shows a strong double-humped light curve.
The reason for this is clear upon reflection: the AM CVn disks are
physically much smaller than the disks in systems with hydrogen-rich
secondary stars, resulting in a much higher specific kinetic energy to
be dissipated at the bright spot since the disk rim is much deeper in
the potential well of the primary.  The smaller disk may also yield a
smaller amplitude for the disk oscillation signal.  In the
hydrogen-rich systems in permanent outburst, the disks are large, the
mass transfer rates are high, and the disk signal dominates, with a
relatively minor contribution from the stream source.  

\begin{figure}
\epsscale{1.0}
\plotone{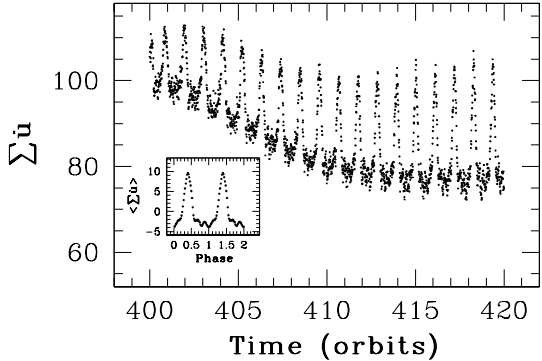}
\caption{The first 20 orbits after SPH simulation restart with no mass
flow through L1.  The pulse shape increases in amplitude as the mean
luminosity increases, resulting from the expansion of the disk.  Note
that the pulse shape is strongly -- and singly -- peaked.  The mean
pulse shape calculated over orbits 410-440 is shown in the inset.  }
\label{fig: en400420ns} 
\end{figure}

We tested the viability of the two-source model through three additional
numerical experiments.  First, we again restarted the above simulation at
orbit 400, but now with the accretion flow through L1 shut off
completely.  In this run, there is no accretion stream and hence no
bright spot contribution.  We show the first 20 orbits of the
simulation light curve in Figure~\ref{fig: en400420ns}.  With the stream
present, the light curve has the double-humped shape of
Figure~\ref{fig: en400420} above, but without the stream the light curve
is sharply peaked with no hint of a double hump. Note that because
there is no low-specific-angular-momentum material accreting at the
edge of the disk, the disk can expand further into the driving zone.
This expansion results in the pulse shape growing in amplitude as the
mean disk luminosity drops.  The pulse shape averaged over orbits
410-440 is shown as an inset in the Figure, and clearly shows that the
oscillating disk is the only source of modulation in the light curve
-- maximum brightness corresponds to a disk geometry like that from
panel 1 of Figure~\ref{fig: sph+} above.  The mean brightness is
roughly constant for orbits 410-440, and at orbit 440 the mean
brightness and pulse amplitude begin to decline as some 50\% of the
initially-present SPH disk particles are accreted by orbit 450. 

\begin{figure}
\epsscale{1.0}
\plotone{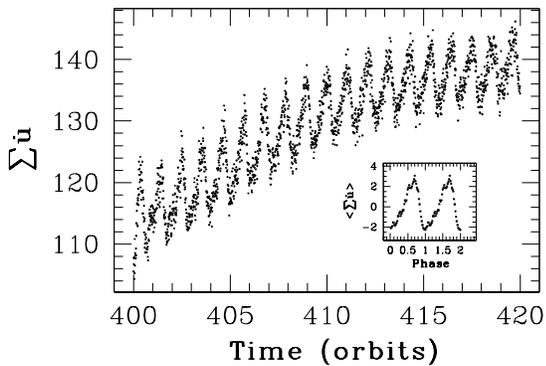}
\caption{The first 20 orbits after SPH simulation restart with
enhanced mass
flow through L1 as described in the text.  The pulse shape is
saw-toothed, and maximum light corresponds to the phase where the
bright spot is deepest in the potential well of the non-axisymmetric
flexing disk. The average pulse shape over these 20 orbits is shown as
an inset.
The mean system luminosity increased with the simulation particle
count.  The mean
pulse shape calculated over orbits 410-440 is shown in the inset.  }
\label{fig: en400420burst} 
\end{figure}

Our second test was to restart the simulation a third time at orbit
400, but this time to enhance the injection rate of SPH particles
(mass flow) at L1 by roughly a factor of 2 over that required to keep
the disk particle count constant (Figure \ref{fig: en400420burst}).
This enhanced mass flux again dramatically changes the character of
the light curve.  Here the mean pulse shape as shown in the inset is
saw-toothed, but with the substructure near the peak from the
interaction of the stream with the periodic motion of the spiral
features in the disk as viewed in the co-rotating frame.  Careful
comparison of the times of maximum in these two runs
(Figures~\ref{fig: en400420ns} and \ref{fig: en400420burst}) reveals
that they are antiphased with each other.  For example, the simulation
light curve in Figure~\ref{fig: en400420ns} shows maxima at times of
403.0 and 404.0 orbits, whereas the simulation light curve in
Figure~\ref{fig: en400420burst} shows minima at these same times.

Our third experiment was more crude, but still effective.  We began
with a disk from a $q=0.2$ low-viscosity SPH simulation run that was
in a stable, non-oscillating state.  We offset all of the the SPH particles an amount
$0.03a$ along the line of centers [i.e., $(x,y,z)\rightarrow (x+0.03a,y,z)$], scaled the SPH particle speeds (but not directions) using the {\it vis viva} equation 
\begin{equation}
v^2 = GM_1\left({{2\over r}-{1\over a}}\right), 
\end{equation} 
and restarted the
simulation.  This technique gives us disk which is non-axisymmetric
but not undergoing the superhump oscillation.  The results were as
expected: we find maxima in the simulation light curves at the phases
where the accretion stream impacts the disk edge deepest in the
potential well of the primary.

In summary, numerical simulations reproduce the two-source model for
positive superhumps.  

\subsection{Negative Superhumps}

\begin{figure}
\epsscale{1.0}
\plotone{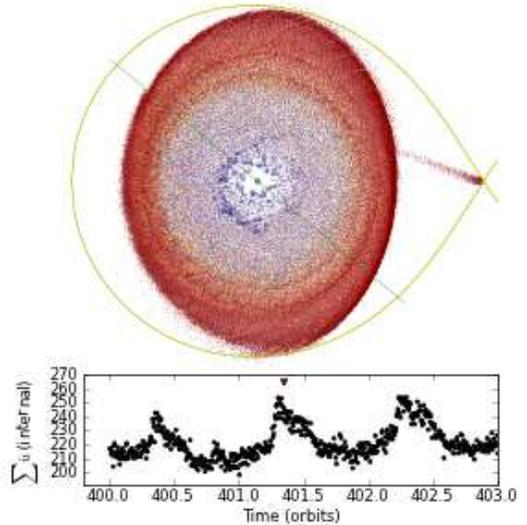}
\caption{A snapshot at orbit 401.35 
from a $q=0.40$ SPH simulation showing negative
superhumps.  The green line running diagonally though the
primary indicates the location of the line of nodes; the disk midplane
includes this line, but is below the orbital plane to the right of the
line, and above the orbital plane to the left of the line.  The particles are color 
coded by luminosity, and the brightest are rendered larger.  It can be seen that the accretion stream impact region at this phase is deep in the potential well of the primary, roughly at the line of nodes. The deeper the accretion stream impact region, the brighter the bright spot.
The simulation light curve is derived from the
``surface'' particles calculated using a simply ray-trace technique.  }
\label{fig: sph-}
\end{figure}

Photometric signals with periods a few percent shorter than $\Porb$
have also been observed in several DN, novalikes, and AM CVn systems
-- in some cases simultaneously with positive superhumps \citep[see,
e.g., Table 2 of][and Woudt et al. 2009]{wts09}.  These oscillations
have been termed {\it negative} superhumps owing to the sign of the
period ``excess'' obtained using Equation~\ref{eq: eps+}.  The system
TV Col was the first system to show this signal, and \citet{bbmm85}
suggested that the periods were consistent with what would be expected
for a disk that was tilted out of the orbital plane and freely
precessing with a period of $\sim$4 d.  \citet{bow88} expanded on this
and suggested what is now the accepted model for the origin of
negative superhumps: the transit of the accretion stream impact point
across the face of a tilted accretion disk that precesses in the
retrograde direction \citep[see][]{wms00,wb07,wts09,foulkes06}.  As in
the stream source for positive superhumps, the modulation results
because the accretion stream impact point has a periodically-varying
depth in the potential well of the primary star.

Finding the term ``negative
period excess'' unnecessarily turgid, in this work we refer to
the {\it  period deficit} $\epsilon_-$ defined as
\begin{equation}
\epsilon_-\equiv {\Porb - P_-\over\Porb}.
\label{eq: eps-}
\end{equation}
Empirically, it is found that for systems showing both positive and
negative superhumps that $\epsilon_+/\epsilon_-\sim2$
\citep{patterson99,retterea02}.

We show in Figure~\ref{fig: sph-} a snapshot from a $q=0.40$
simulation that demonstrates the physical origin of negative
superhumps.  At orbit 400, the disk particles were tilted $5^\circ$
about the $x$-axis and the simulation restarted.  The green line in
the Figure running diagonally though the primary indicates the
location of the line of nodes; the disk midplane includes this line,
but is below the orbital plane to the right of the line, and above the
orbital plane to the left of the line.  The disk particles are again
color-coded by luminosity, and the brightest particles are shown with
larger symbols.  The ballistic accretion stream can be followed from the L1
point to the impact point near the line of nodes.  The simulation
light curve is derived from the ``surface'' particles as described in
\citet{wb07}.  The times of maximum of the negative superhump light
curve occur when accretion stream impact point is deepest in the
potential of the primary and on the side of the disk facing the
observer.  A second observer viewing the disk from the opposite side
would still see negative superhumps, but antiphased to those of the
first.

Having  introduced a viable model for positive superhumps and their
evolution, let us now compare the model to the \Kepler V344 Lyr
photometry.

\section{\Kepler Photometric Observations}

The primary science mission of the NASA Discovery mission \Kepler is
to discover and characterize terrestrial planets in the habitable zone
of Sun-like stars using the transit method \citep{borucki10,haas10}.
The spacecraft is in an Earth-trailing orbit, allowing it to view its
roughly 150,000 target stars continuously for the 3.5-yr mission
lifetime.  The photometer has no shutter and stares continuously at
the target field.  Each integration lasts 6.54 s.  Due to memory and
bandwidth constraints, only data from the pre-selected target
apertures are kept.  \Kepler can observe up to 170,000 targets using
the long-cadence (LC) mode, summing 270 integrations over 29.4 min,
and up to 512 targets in the short-cadence (SC) mode, summing 9
integrations for an effective exposure time of 58.8 s.  

\begin{figure*}
\plotone{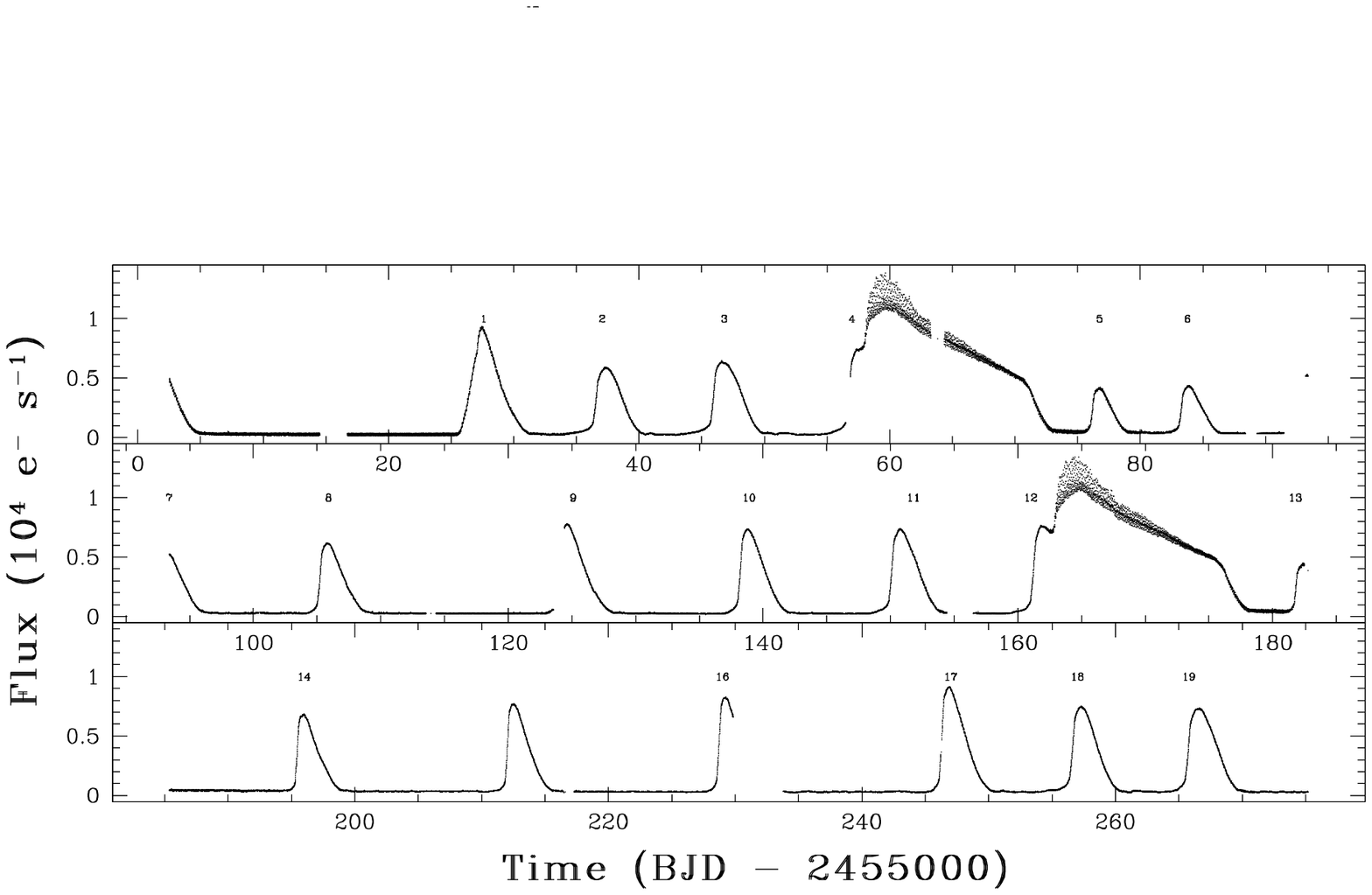}
\caption{The \Kepler Q2, Q3, and Q4 light curve of V344 Lyr in flux
units.  The outbursts are labeled 1 to 19.  
}
\label{fig: lcrawflux3}
\end{figure*}

There are gaps in the \Kepler data streams resulting from, for
example, monthly data downloads using the high-gain antenna and
quarterly 90$^\circ$ spacecraft rolls, as well as unplanned safe-mode
and loss of fine point events.   For further details of the spacecraft
commissioning, target tables, data collection and processing, and
performance metrics, see \citet{haas10}, \citet{koch10}, and
\citet{caldwell10}.  

\Kepler data are provided as quarterly FITS files by the Science
Operations Center after being processed through the standard data
reduction pipeline \citep{jenkins10}.  The raw data are first
corrected for bias, smear induced by the shutterless readout, and sky
background.  Time series are extracted using simple aperture
photometry (SAP) using an optimal aperture for each star, and these
``SAP light curves'' are what we use in this study.  The dates and
times for the beginning and end of Q2, Q3 and Q4 are listed in
Table~\ref{tbl: quarters}.

\begin{deluxetable*}{ccccc}
\tablewidth{0pt}
\tablecaption{Journal of Observations}
\tablehead{
\colhead{Quarter} & \multicolumn{2}{c}{Start\tablenotemark{a}} & \multicolumn{2}{c}{End\tablenotemark{b}} \\
\colhead{\ } & \colhead{MJD} & \colhead{UT} &  \colhead{MJD}   & \colhead{UT} }
\startdata
Q2 & 55002.008 & 2009 Jun 20 00:11 & 55090.975 & 2009 Sep 17 11:26 \\
Q3 & 55092.712 & 2009 Sep 18 17:05 & 55182.007 & 2009 Dec 17 00:09 \\
Q4 & 55184.868 & 2009 Dec 19 20:49 & 55274.714 & 2010 Mar 19 17:07
\enddata
\tablenotetext{a}{The start MJD and UT dates are the mid-point of the first cadence of the SC time series for each quarter.}
\tablenotetext{b}{The end MJD and UT dates are the mid-point of the last cadence of the SC time series for each quarter.}
\label{tbl: quarters}
\end{deluxetable*}

The full SAP light curve for \Kepler quarters Q2, Q3, and Q4 is shown
in flux units in Figure~\ref{fig: lcrawflux3}.  In Figure 2 of Paper
II we show the full SAP light curve in Kp magnitude units.  As noted
in Paper II and evident in Figure~\ref{fig: lcrawflux3}, the
superoutbursts begin as normal DN outbursts.

The Q2 data begin at BJD   2455002.5098.  For simplicity we will
below refer to events as occurring on, for example, day 70, which
should be interpreted to mean BJD 2455070 -- that is we take BJD
2455000 to be our fiducial time reference.

In this paper, we focus on the superhump and orbital signals present
in the data.  The outburst behavior of these data in the context of
constraining the thermal-viscous limit cycle is published separately
(Paper II).  

\begin{figure*}
\plotone{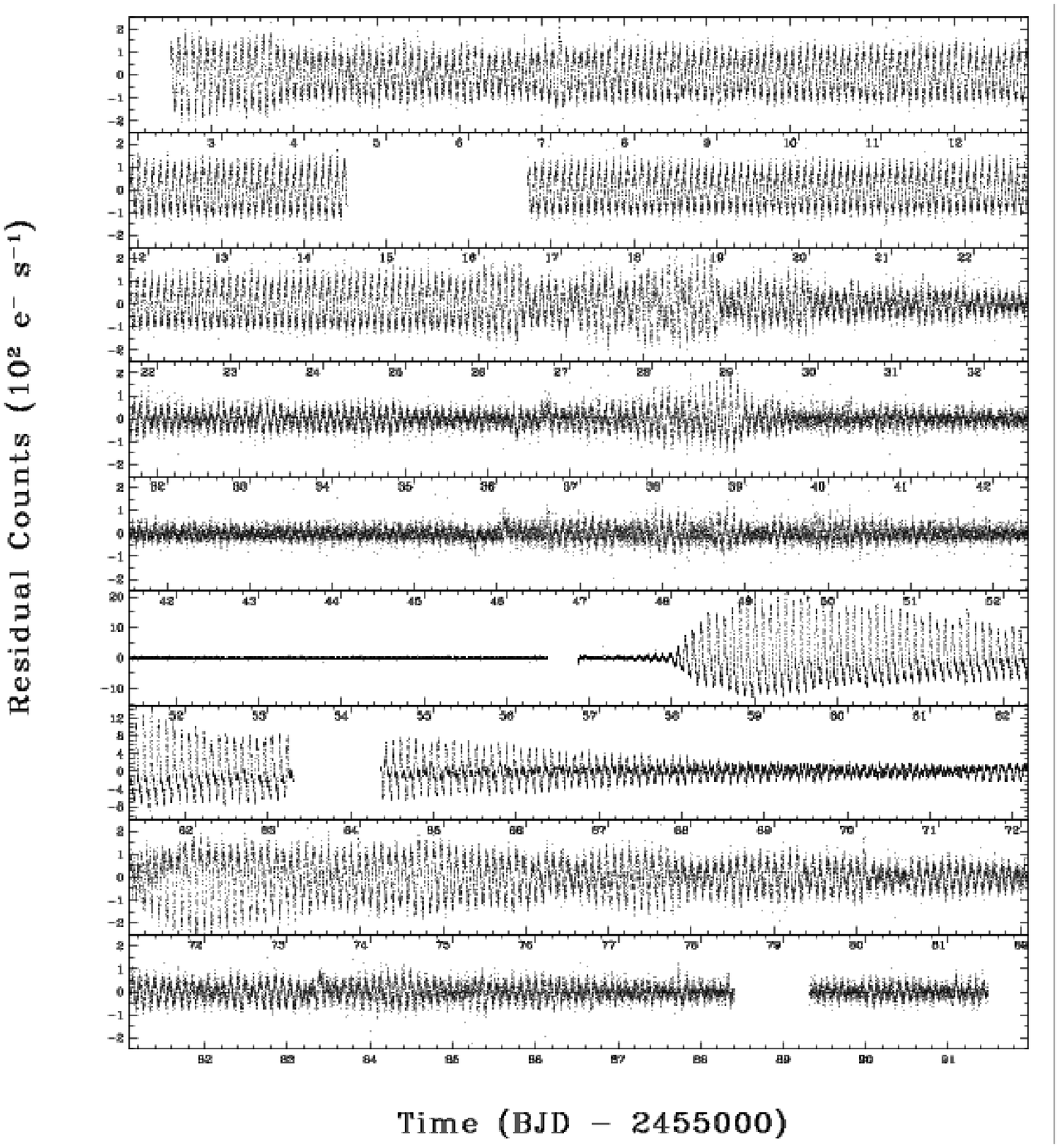}
\caption{The residuals of the \Kepler Q2 V344 Lyr light curve 
after subtraction of a smoothed copy of the SAP light curve.  
Each panel shows approximately 10 days of data ($\sim$120 cycles).
Note that the vertical scaling of the 2 panels showing the
high-amplitude positive superhumps ($P_+ = 2.20$ hr, $\nu_+ =  10.9\cd$) differ from the rest (days $\sim$52-72), and that there is a small amount of overlap between panels.  The negative superhump signal dominates in days $\sim$2--40 ($P_- = 2.06$ hr period, $\nu_- = 11.7\cd$).
The positive
superhumps  dominate the power
for days $\sim$58 through the end of the quarter.
}
\label{fig: reslc1}
\end{figure*}

\begin{figure*}
\plotone{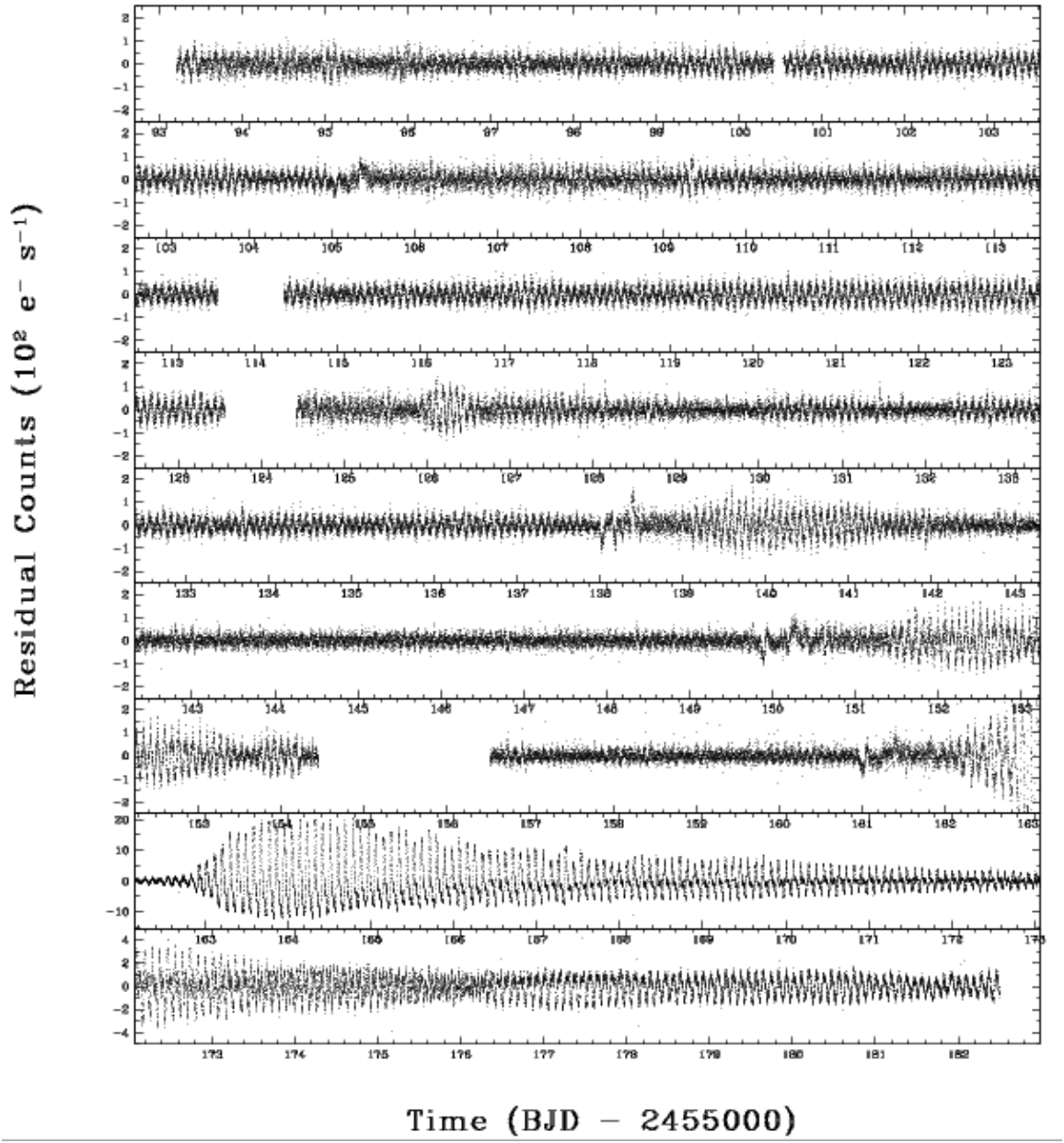}
\caption{The residuals of the \Kepler Q3 V344 Lyr light curve. 
The positive superhumps dominate the
power for days $\sim$162 through the end of the quarter.
The negative superhumps dominate the signal during days $\sim$100-160. The vertical scaling of the bottom 2 panels differs from the rest.
The features that appear during days $\sim$105, 138, 150, and 161 are
artifacts of the reduction process.
}
\label{fig: reslc2}
\end{figure*}

\begin{figure*}
\plotone{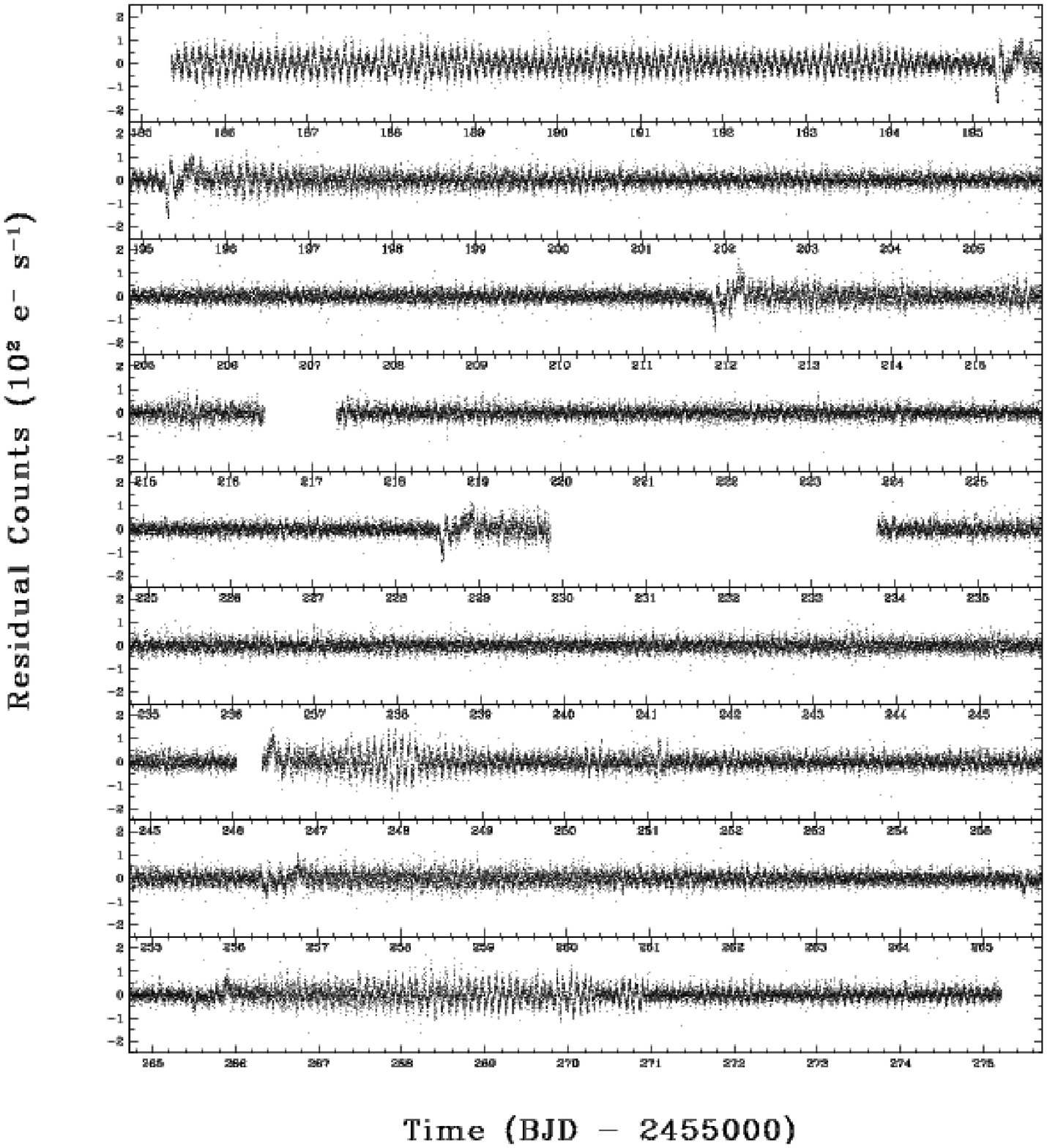}
\caption{The residuals of the \Kepler Q4 V344 Lyr light curve.  The orbital signal ($\Porb = 2.11$ hr, $\nuorb = 11.4\cd$) dominates following the decay of the postive superhump signal at day $\sim$190.
The features that appear during days $\sim$195, 212, 228, 246, and 256
are artifacts of the reduction process.
}
\label{fig: reslc3}
\end{figure*}

To remove the large-amplitude outburst behavior from the raw light
curve -- i.e., to high-pass filter the data -- we subtracted a
boxcar-smoothed copy of the light curve from the SAP light curve.  The
window width was taken to be the superhump cycle length (2.2 hr or 135
points).  To minimize the effects of data gaps, we split the data into
a separate file anytime we had a data gap of more than 1 cycle.  This
resulted in 10 data chunks.  Once the data residual light curve was
calculated, we again recombined the data into a single file.  The
results for Q2, Q3, and Q4 are shown in Figures~\ref{fig: reslc1},
\ref{fig: reslc2}, and \ref{fig: reslc3}, respectively. 

We also calculated the fractional amplitude light curve by dividing
the raw light curve by the smoothed light curve, and subtracting 1.0.
However, as expected, the amplitudes of the photometric signals in the
residual light curve are more nearly constant than those in the
fractional amplitude light curve.  This is because the superhump
signals -- both positive and negative -- have amplitudes determined by
physical processes within the disk that are not strong functions of
the overall disk luminosity.

\section{The Fourier Transform}

\begin{figure*}
\epsscale{1.0}
\plotone{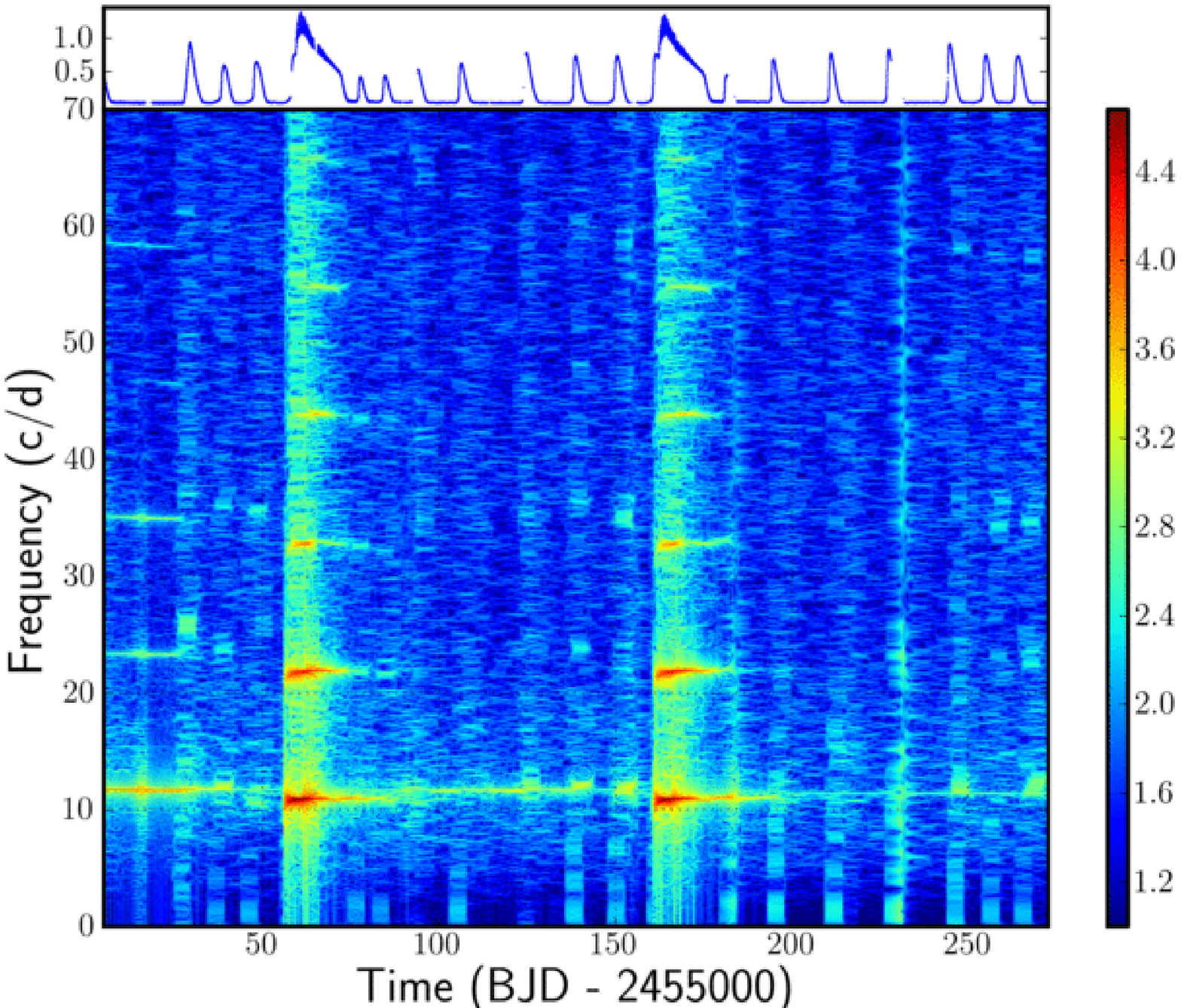}
\caption{The 2D DFT of the Kepler Data from Q2, Q3, and Q4.  Shown
here is the logarithm of the amplitude versus time and frequency.
There are three fundamental periods visible, plus higher harmonics of two of
these. 
The positive superhumps ($P_+ = 2.20$ hr, $\nu_+ =  10.9\cd$) dominate the
power for days $\sim$58--80 and $\sim$162--190.
The negative superhumps dominate the signal during days $\sim$100--160 ($P_- = 2.06$ hr, $\nu_- = 11.7\cd$) dominate early in Q2, and again in Q3 between the superoutbursts.
The orbital signal ($\Porb = 2.11$ hr, $\nuorb = 11.4\cd$) is apparent during Q4, and once identified, during the week before the second superoutburst.}
\label{fig: 2dDFT}
\end{figure*}

\begin{figure*}
\epsscale{1.0}
\plotone{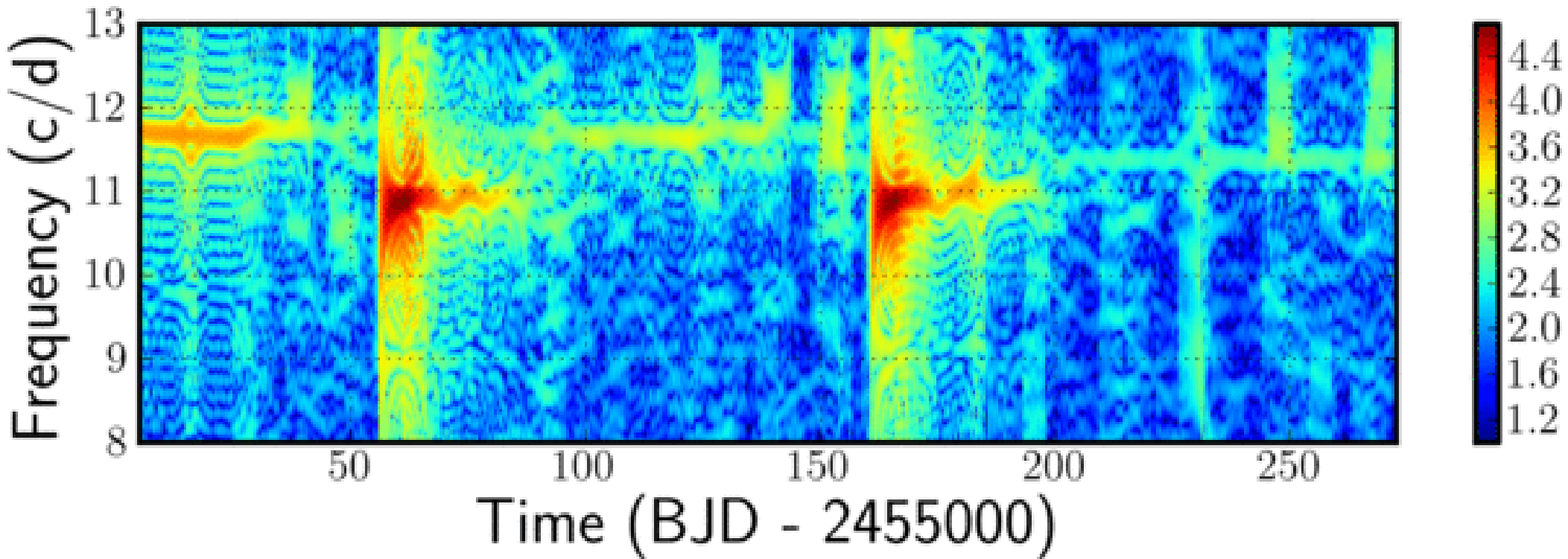}
\caption{Magnified view of the 2D DFT
for quarters Q2, Q3, and Q4.  Shown here is the logarithm of the
amplitude versus time and frequency.  The three distinct
frequencies are clearly evident.  In days $\sim$2--40 and again in
days $\sim$100--160, the $P_- = 2.06$ hr period ($\nu_- = 11.7\cd$)
indicative of the negative superhump dominates.  The positive
superhumps ($P_+ = 2.20$ hr, $\nu_+ =  10.9\cd$) dominate the power
for days $\sim$58--80 and $\sim$162--190.  The orbital period ($\Porb
= 2.11$ hr, $\nuorb = 11.4\cd$) is most clearly apparent in the Q4 data,
starting about day 200.}
\label{fig: 2dDFTzoom}
\end{figure*}

In Figure~\ref{fig: 2dDFT} we show the discrete Fourier transform
amplitude spectra for the current data set.  We took the transforms
over 2000 frequency points spanning 0 to 70 cycles per day.  Each
transform is of a 5 day window of the data, and the window was moved
roughly 1/2 day between subsequent transforms.  The color scale
indicates the logarithm of the residual count light curve amplitude in
units of counts per cadence.  In Figure~\ref{fig: 2dDFTzoom} we show a
magnified view including only frequencies 9.5 to 12.5 c/d to better
bring out the 3 fundamental frequencies in the system.

Figures~\ref{fig: 2dDFT} and \ref{fig: 2dDFTzoom} are rich with
information.  The positive superhumps ($P_+ = 2.20$ hr) dominate the
power for days $\sim$58--80 and $\sim$162--190.  In Figure~\ref{fig:
2dDFTzoom} we see the that time evolution of the fundamental
oscillation frequency is remarkably similar in both superoutbursts.
The dynamics behind this are discussed below in \S5.2 where the O-C
diagrams are presented.

\begin{figure} \plotone{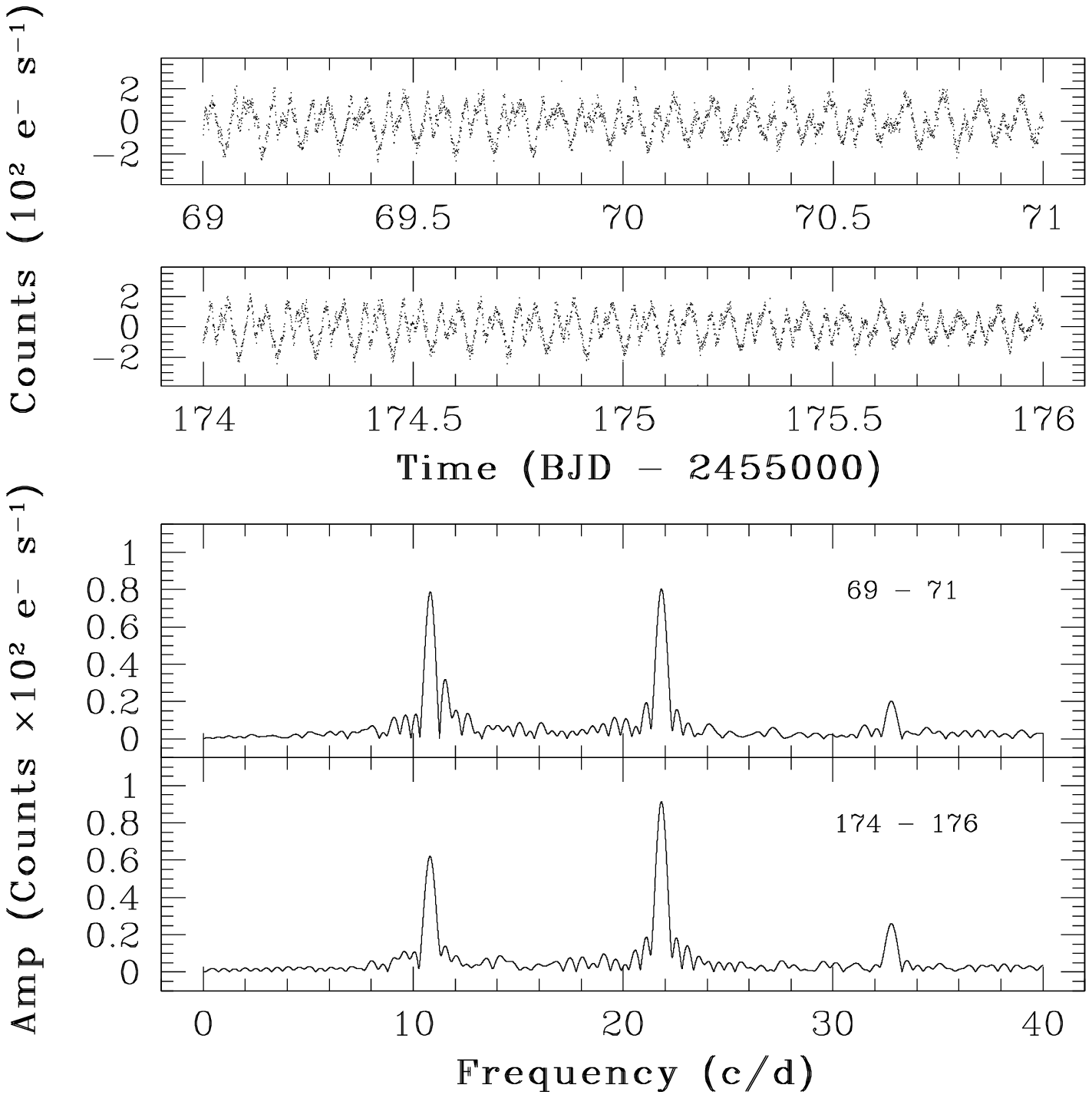}
\caption{Light curves and associated Fourier transforms during the transition from disk to stream superhumps. As the system transitions from disk to stream superhumps in the drop to quiescence, the light curve is strongly double peaked but non-sinusoidal, resulting in power at both $\nu_+$ and $2\nu_+$ (first and second harmonics).}
\label{fig: trans}
\end{figure}

Once the majority of the mass that will accrete during the event has
done so, the disk transitions back to the low state.  This occurs
roughly 15 d after superhump onset for V344 Lyr.  During this
transition, the disk source of the superhump modulation fades with the
disk itself, and the stream source of the superhump modulation begins
to dominate.  A careful inspection of Figure~\ref{fig: 2dDFT} shows
that at this time of transition between disk and stream superhumps,
there is comparable power in the second harmonic (first overtone) 
as found in the fundamental.  The behavior of the light curve and
Fourier transform are more clearly displayed in Figure~\ref{fig:
trans} which shows 2 days of the light curve during the transition
period, and the associated Fourier transforms.  In both cases, the
``knee'' in the superoutburst light curve (see Figure~\ref{fig:
lcrawflux3} occurs just past the midpoint of the data sets.  Although
the second harmonic is strong in transition phase, the pulse shape of
the disk superhump signal is sharply peaked so the fundamental remains
prominent in the Fourier transform (see Figure~\ref{fig: trans}).  

As can clearly be seen in Figure~\ref{fig: 2dDFTzoom}, the orbital
period of $2.10$ hr (11.4 c/d) only becomes readily apparent in the Q4
data, starting at about day 200, and it dominates the Q4 Fourier
transforms.  Once identified in Q4, the orbital frequency appears to
show some power in the week before the first superoutburst in Q2, and
between days $\sim$130 and the second superoutburst in Q3.   Note,
however, that the amplitude of the orbital signal is roughly 1 order
of magnitude smaller than the amplitude of the negative superhump
signal, and as much as 2 orders of magnitude smaller than the
amplitude of the positive superhump signal.  In these data, the
orbital signal is found only when the positive or negative superhump
signals are weak or absent.  We discuss the physical reason for this
below.

Finally, we note that we searched the Fourier transform of our \Kepler
short-cadence (SC) data out to the Nyquist frequency of 8.496 mHz for
any significant high frequency power which might for example indicate
accretion onto a spinning magnetic primary star (i.e., intermediate
polar or DQ Her behavior).  We found no reliable detection of higher
frequencies in the data, beyond the well-known spurious frequencies
present in \Kepler time series data at multiples of the LC frequency
\citep[$n\times0.566427$ mHz $=$ 48.9393 $\rm c\
d^{-1}$][]{gilliland10}.  For a full list of possible spurious
frequencies in the SC data, see the {\it Kepler Data Characteristics
Handbook}.

\subsection{The Orbital Period}

\begin{figure}
\plotone{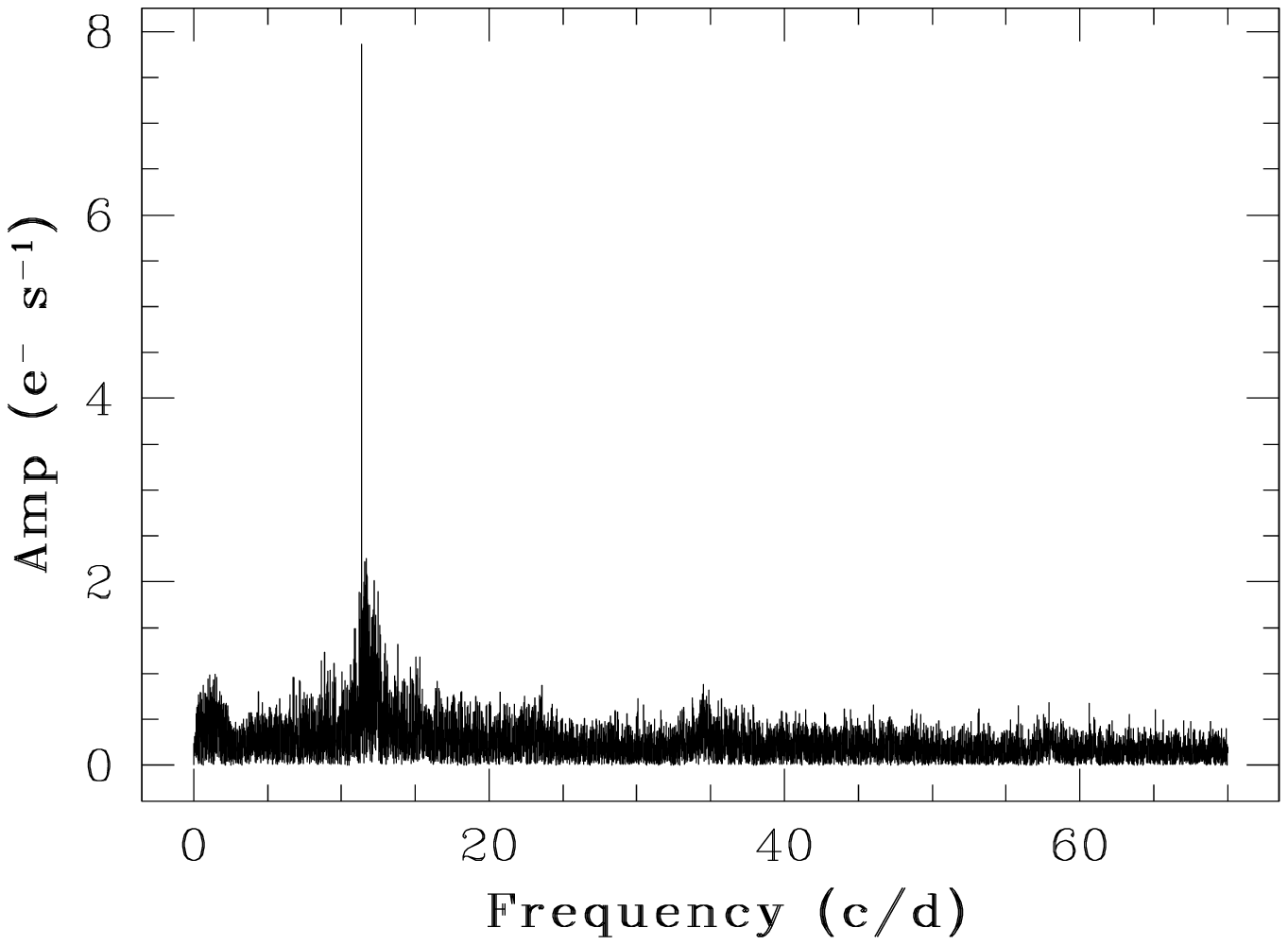}
\caption{The Fourier transform obtained using the Q4 residual data.
The single dominant peak is at 11.38 c/d (2.11 hr).   We identify this
as the orbital period.}
\label{fig: q4dft}
\end{figure}

\begin{figure}
\epsscale{1.0}
\plotone{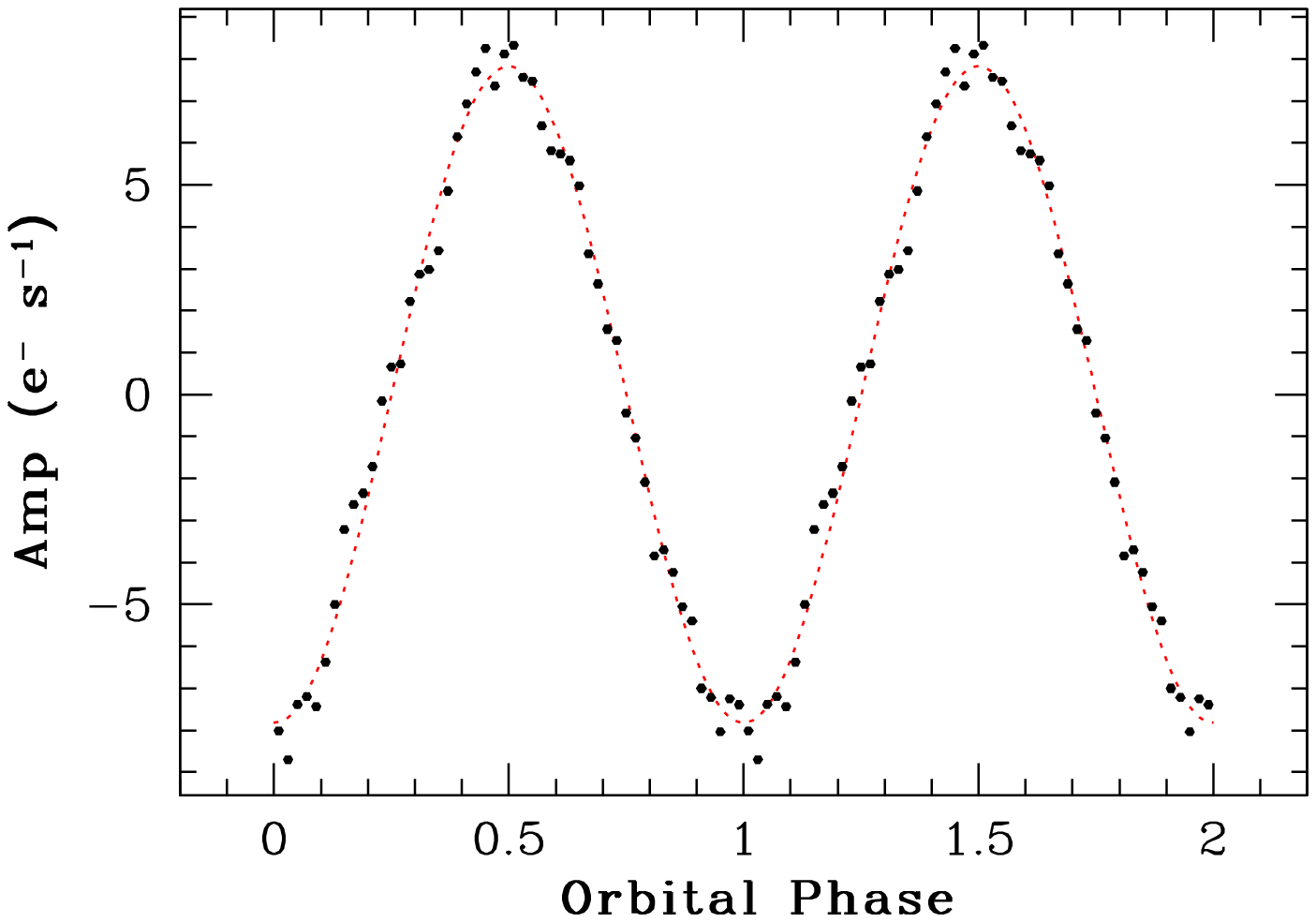}
\caption{The average pulse shape for the identified orbital period,
averaged over days 200-275 in Q4.  Also shown for comparison is a sine
curve.}
\label{fig: avelcporb}
\end{figure}

The orbital period is the most fundamental clock in a binary system.
In the original Q2 data presented by \citet{still10}, the only
frequencies that were clearly present in the data were the 2.20-hr
(10.9 c/d) superhump period and the period observed at 2.06-hr (11.7
c/d).  In Paper I we identified this latter signal as the orbital
period but discussed the possibility that it is a negative superhump
period.  The Q3 data revealed a marginal detection of a period of 2.11
hr (11.4 c/d), and this period is found to dominate the Q4 data (see
Figure~\ref{fig: q4dft}).  The average pulse shape for this signal
averaged over days 200-275 is shown in Figure~\ref{fig: avelcporb}.
We can now safely identify this 2.11 hr (11.4 c/d) signal as the
system orbital period, which then indicates that the 2.06 hr (11.7
c/d) signal is a negative superhump.

The orbital period was determined using the method of non-linear least
squares fitting a function of the form
\begin{equation}
y(t) = A \sin[2\pi(t-T_0)/P].
\end{equation}
The results of the fit are
\begin{eqnarray*}
P &=& 0.087904\pm3\times10^{-6}\rm\ d,\\
  &=& 2.109696\pm7\times10^{-5}\rm\ hr,\\
T_0 &=& {\rm BJD}\ 2455200.2080\pm0.0006,\\
A &=& 7.8\pm 0.1\rm\ e^-\ s^{-1}.
\end{eqnarray*}
Note that the amplitude is only roughly 25 mmag  -- an order of
magnitude or more smaller than the peak amplitudes of the positive and
negative superhumps in the system.  

That an orbital signal exists indicates that the system is not
face-on.  The source of the orbital signal of a non-superhumping
CV can be either the variable flux along the line of site from a
bright spot that is periodically shadowed as it sweeps around the back
rim of the disk, or the so-called reflection effect as the face of the
secondary star that is illuminated by the UV radiation of the disk
rotates in to and out of view \citep[e.g.,][]{warner95}.  In
Figure~\ref{fig: 2dDFTzoom}, we find that the orbital signal is never
observed when the positive superhumps are present, but this is not a
strong constraint as the positive superhump amplitude swamps that of
the orbital signal.  

More revealing is the interplay between the orbital signal, the
negative superhump signal, and the DN outbursts.  In Q2 and Q3, the
orbital signal appears only when the negative superhump signal is weak
or absent.  This is consistent with the idea that  the addition of
material from the accretion stream should bring the disk back to the
orbital plane roughly on the mass-replacement time scale
\citep{wb07,wts09}.  The strong negative superhump signal early in Q2
indicates a tilt of $\sim$5$^\circ$, sufficient for the accretion
stream to avoid interaction with the disk rim for all phases except
those in which the disk rim is along the line of nodes.  As the disk
tilt declines, however, an increasing fraction of the stream material
will impact the disk rim and not the inner disk -- in other words, the
orbital signal will grow at the expense of the negative superhump
signal.  This appears to be consistent with the data in hand and if so
would suggest that the orbital signal results from the bright spot in
V344 Lyr, but the result is only speculative at present.  

\begin{figure}
\epsscale{1.0}
\plotone{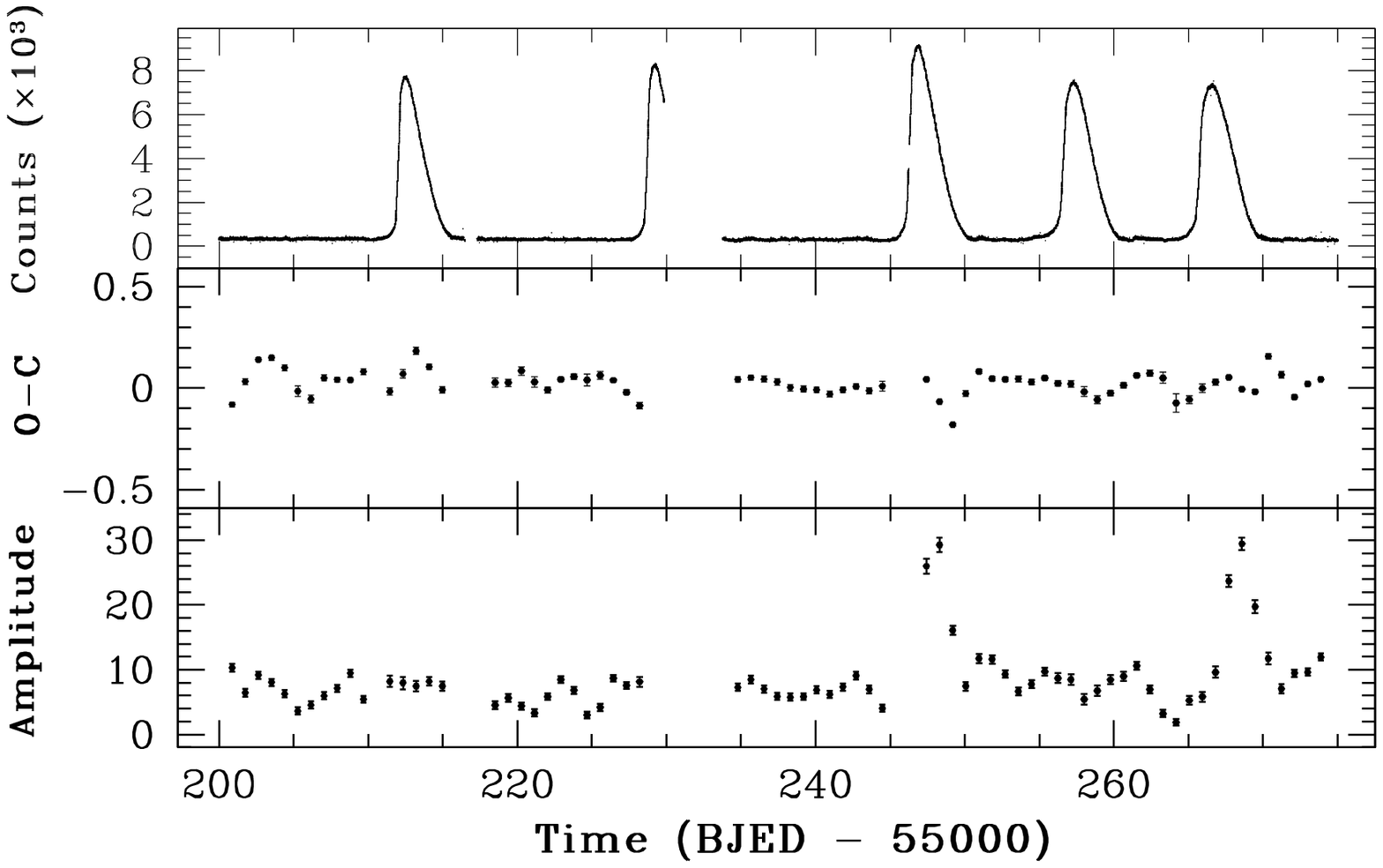}
\caption{The O-C phase diagram for the identified orbital period
$\Porb=2.10$ hr.  For each point 20 cycles were fit, and each window
is offset from the previous by 10 cycles.  The bottom panel shows the
amplitude of the fits to the residual light curves.  The phase is
steady, showing only small variations resulting from interference with
other periods that are present, and that correlate with times of
outburst.  The large amplitudes found during outbursts 17 and 19
result from the higher-frequency signal initiated on decline from
maximum in each case (see Figure~\ref{fig:  2dDFTq4}).
}
\label{fig: omc200275}
\end{figure}

\begin{figure*}
\plotone{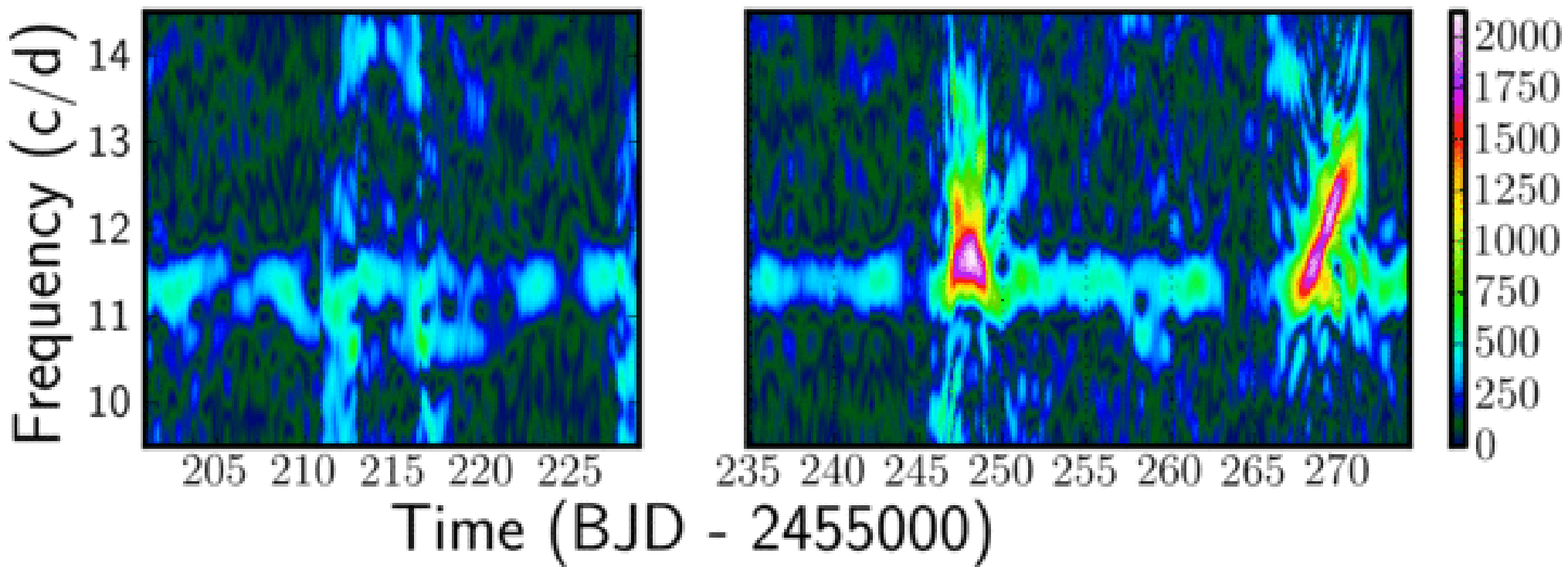}
\caption{The 2D DFT of the post-superoutburst Q4 data.  Shown are
amplitude per cadence versus time and frequency.  The window
width was 2 days, and the window was shifted  1/8 day between
transforms. The orbital signal is present but weak for most of the
quarter.  Power at a higher frequency is found on the decline from
maximum in outbursts 17 and 19. }
\label{fig: 2dDFTq4}
\end{figure*}

In Figure~\ref{fig: omc200275} we show the O-C phase diagram for
$\Porb$.  We fit 20 cycles for each point in the Figure, and
moved the window 10 cycles between fits.  The small apparent
wanderings in phase result from interference from the other periods
present, and also appear to correlate with the outbursts.    We show
the 2D DFT for days 200 to 275 in Figure~\ref{fig: 2dDFTq4}.  Here we
used a window width of 2 days, and shifted the window by 1/8th of a
day between transforms.  We show amplitude per cadence.
The orbital signal appears to be increasing in amplitude slightly
during Q4, perhaps as a result of the buildup of mass in the outer
disk after several DN outbursts.  The large amplitudes found for the
orbital signal in Figure~\ref{fig: omc200275} during outbursts 17 and 19
(starting days $\sim$246.5 and 266, respectively) are spurious,
resulting from the higher-frequency signals found on the decline from
maximum in each case.  As discussed below, outbursts 17 and 19 both show
evidence for triggering a negative superhump signal, and the light
curve for outburst 19 yields a complex Fourier transform that shows power
at the orbital frequency, the negative superhump frequency, and at
12.3 c/d (1.95 hr).  

\subsection{Observed Positive Superhumps}

The light curve for V344 Lyr is rich in detail, and in particular
provides the best data yet for exploring the time evolution of
positive superhumps.  As discussed above, the superhumps are first
driven to resonance during the DN outburst that precedes the
superoutburst as the heating wave transitions the outer disk to the
high-viscosity state allowing the resonance to be driven to
amplitudes that can modulate the system luminosity.  
Close inspection of the positive
superhumps in Figures~\ref{fig: reslc1} and~\ref{fig: reslc2} shows
that in both cases the amplitude of the superhump is initially quite
small, but grows to saturation ($A\sim0.25$ mag) in roughly 16 cycles.
There is a signal evident preceding the second superoutburst (days
$\sim$156.5 to 161) -- this is a blend of the orbital signal and a
very weak negative superhump signal.  

The mean superhump period obtained by averaging the results from
non-linear least squares fits to the disk superhump signal during the
two superoutburst growth through plateau phases is $P_+ = 0.091769(3)\rm\ d =
2.20245(8) hr$.  The errors quoted for the last significant digit are
the {\it formal} errors from the fits summed in quadrature.  The
periods drift significantly during a superoutburst, however,
indicating these formal error estimates should not be taken seriously.
Using the periods found for the superhumps and orbit, we find a period
excess of $\epsilon_+ = 4.4\%$.  We plot the result for V344 Lyr with
the results from the well-determined systems below the period gap
listed in Table 9 of \citet{patterson05} in Figure~\ref{fig:
epsvporb}.  The period excess for V344 Lyr is consistent with the
existing data.

\begin{figure}
\plotone{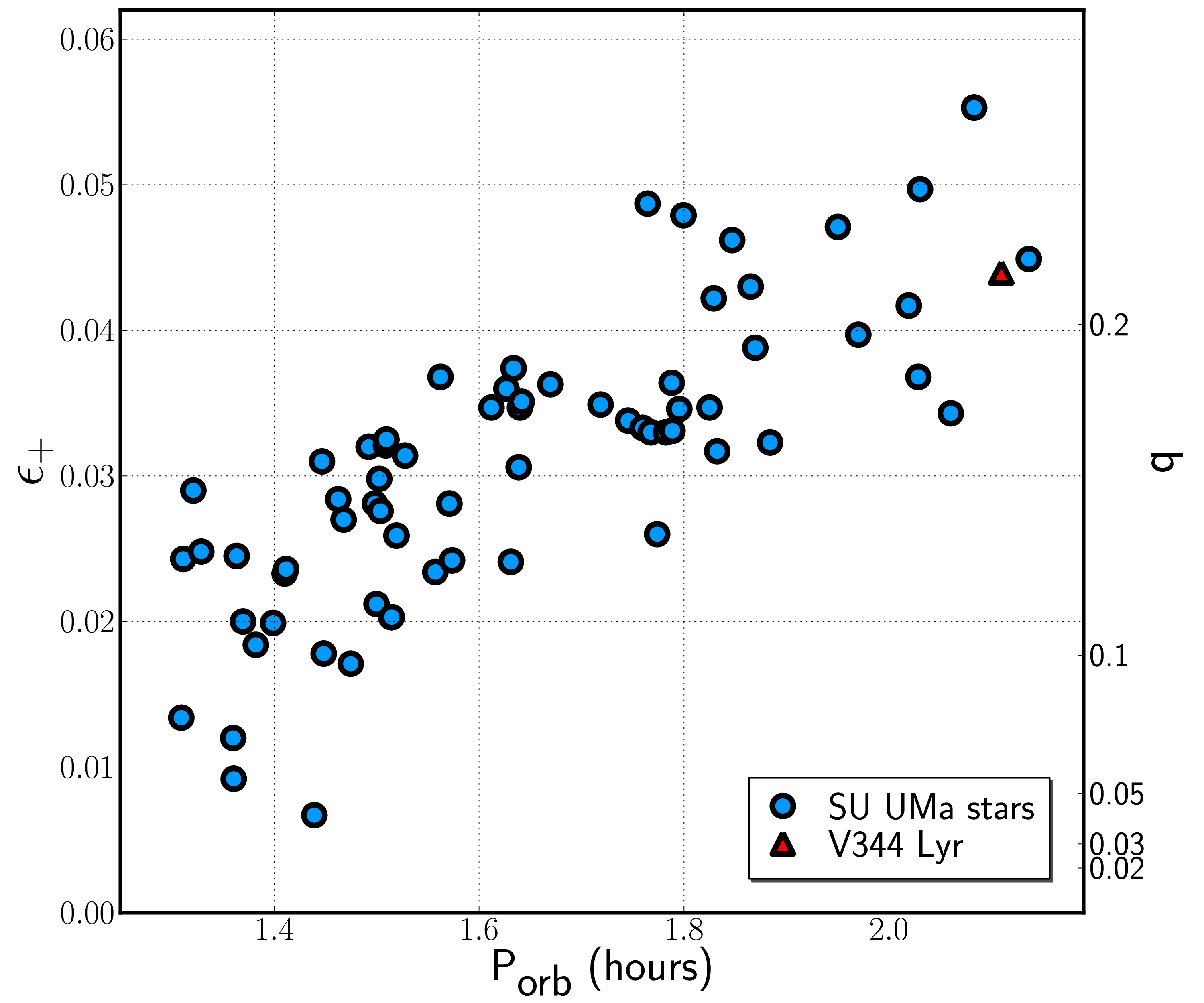}
\caption{Measured period excess versus $\Porb$ data from 
\citep{patterson05} (circles) and for V344 Lyr (triangle).  The system
V344 Lyr is consistent with the trend.}
\label{fig: epsvporb}
\end{figure}

\begin{figure}
\plotone{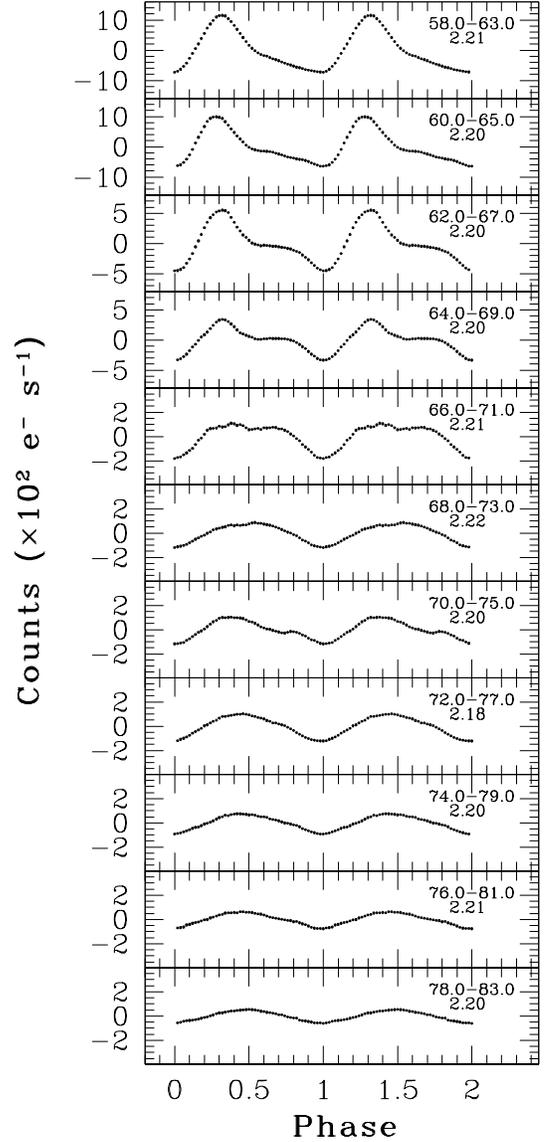}
\caption{The average positive superhump pulse shapes obtained during
the first superoutburst.  The time spans of each 5-day subset of data
and the folding periods (in hr) are indicated in each subpanel.  Zero
phase is set at the primary minimum. The mean pulse shape evolves
considerably over the superoutburst.  Note the changing vertical scale
in the top-few panels.  } 
\label{fig: sh1panave} 
\end{figure}

\begin{figure}
\plotone{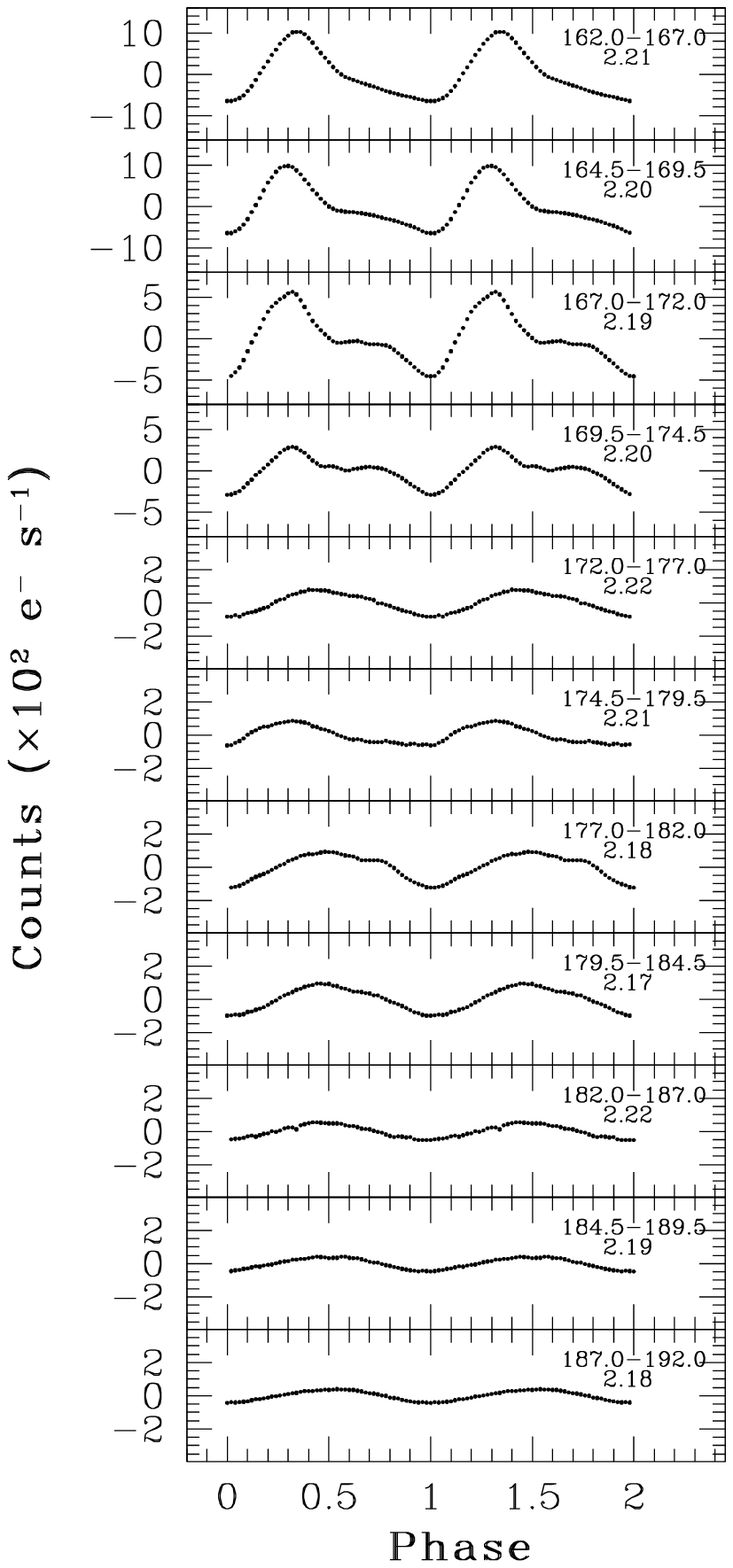}
\caption{Similar to Figure~\ref{fig: sh1panave}, but for the second
superoutburst.  What is particularly interesting is that the evolution
of the pulse shape is remarkably similar to that seen in the first
superoutburst, suggesting that the dynamical evolution is also similar
from one outburst to the next.  } 
\label{fig: sh2panave} 
\end{figure}

In Figures \ref{fig: sh1panave} and \ref{fig: sh2panave} we show the
time evolution of the mean pulse shape for the first and second
superoutbursts.  To create these Figures, we split the data into 5-day
subsets ($\sim$50 cycles), with an overlap of roughly 2.5 days from
one subset to the next.  For each subset we computed a discreet
Fourier transform and then folded the data on the period with the most
power.

The evolution of the mean pulse shape is similar to results published
previously \citep[e.g.][]{patterson03,kato09,kato10}, however the
quality of the \Kepler\ data is such that we can test vigorously 
the model that has been slowly emerging in the past few years
for the origin of the superhump light source, the evolution of the
pulse shape and the physical origin of late superhumps.

A comparison of the simulation light curve from Figure~\ref{fig: sph+}
with the early mean pulse shapes shown in Figures~\ref{fig: sh1panave}
and \ref{fig: sh2panave} reveals a remarkable similarity, all the more
remarkable given the very approximate nature of the artificial
viscosity prescription used in the SPH calculations and the crude way
in which the simulation light curves are calculated.  

If the comparison between data and model is correct, the SPH
simulations illuminate the evolution
of the positive superhumps from the early
disk-dominated source to the late stream-dominated source.  The signal
observed early in the superoutburst is dominated by disk superhumps,
where the disk at resonance is driven into a large-amplitude
oscillation, and viscous dissipation in the strongly convergent flows
that occur once per superhump cycle yield the characteristic
large-amplitude superhumps seen in the top panels of Figures~\ref{fig:
sh1panave} and \ref{fig: sh2panave}.  After $\sim$100 cycles ($\sim$10
d), a significant amount of mass has drained from the disk, and in
particular from the driving region.  The disk continues to oscillate
in response to the driving even after it has transitioned back to the
quiescent state, but the driving is off-resonance and the periodic
viscous dissipation described above is much reduced.  Thus, we agree
with previous authors that the late/quiescent superhumps that have
been observed result from the dissipation in the bright spot as it
sweeps around the rim of the non-axisymmetric disk.

\begin{figure}
\epsscale{0.9}
\plotone{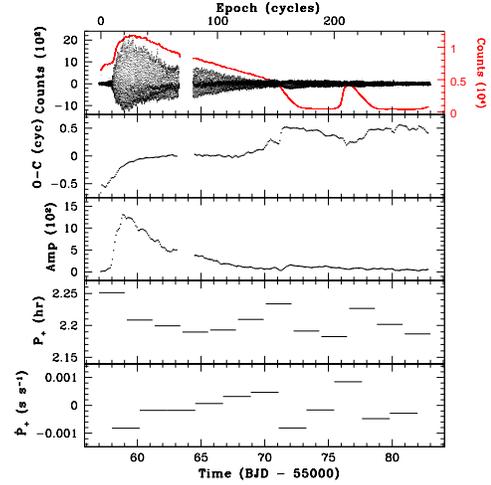}
\caption{The 
 residual light curve (black) in units of $10^2\ e^-\rm\ s^{-1}$ and smoothed
SAP light curve (red) in units of $10^4\ e^-\rm\ s^{-1}$ ({\it top
panel}), O-C phase diagram ({\it second panel}), amplitudes ({\it
third panel}), period $P_+$ ({\it fourth panel}) and localized rates of period
change  $\dot P_+$ ({\it bottom panel}) for the common superhumps present
during the first superoutburst.  The observed phase variations
reflect real variations in the oscillation period as well as the
transition in the dominant modulation source for the light curve.  The
Epoch was arbitrarily set to zero for the first datum in this subset
of the time series.
}
\label{fig: sh1omc}
\end{figure}

\begin{figure}
\epsscale{0.9}
\plotone{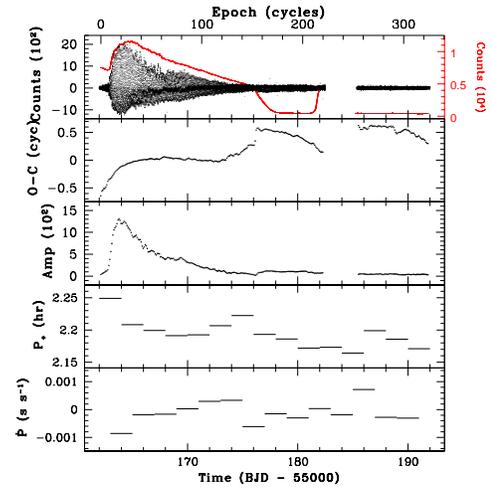}
\caption{Similar to Figure~\ref{fig: sh1omc}, but for the second
superoutburst. This Figure clearly shows that the phase variations
have the same character as before, but with slightly larger excursions from the
mean.
}
\label{fig: sh2omc}
\end{figure}

To compute O-C phase diagrams for each superoutburst,  we fit a
3-cycle sine curve with the mean period of 2.196 hr which yields a
relatively constant O-C during the plateau phase.  The results are
shown in Figures~\ref{fig: sh1omc} and \ref{fig: sh2omc}.  The top
panel shows the residual light curve as well as the SAP light curve
smoothed with a window width of $P_+$ (135 points).  The second panel
shows the O-C phase diagram, and the third panel the amplitude of the
fit.  Also included in this Figure in the fourth panel
are the periods of the positive superhumps during 2-day subsets of
the residual light curve obtained with Fourier transforms.  The
horizontal bars show the extent of each data window.  By differencing
adjacent periods, we calculate the localized rate of period change of
the superhumps $\dot P_+$.  These results are shown in the bottom
panel.  As perhaps might be expected from the similarity in the
evolution of the mean pulse profile during the two superoutbursts, the
O-C phase diagrams as well as the evolution of the periods and
localized rates of period change are also similar.   Such diagrams can
be illuminating in the study of superhumps, and \citet{kato09} and
\citet{kato10} present a comprehensive population analysis of
superhumps using this method.

When the disk is first driven to oscillation in the growth and
saturation phase, there is maximum mass at large radius, and the
corresponding superhump period ($\sim$2.25 hr) is significantly longer
than the mean, yielding a positive slope in the O-C diagram.  The rate
of period change estimated from the first 4 days of data for both
superoutbursts is  $\dot P_+ = -8\times 10^{-4}\ \rm s\ s^{-1}$.
Roughly 10 cycles ($\sim1$ d for V344 Lyr) after the mode saturates
with maximum amplitude, sufficient mass has drained from the outer
disk that the superhump period has decreased to the mean, and the
superhump period continues to decrease out to $E\sim100$ as the
precession rate slows as a result of the decreasing mean radius of the
flexing, non-axisymmetric disk.  The period at this time is roughly
2.19 hr for both superoutbursts, and the rate of period change between
cycles 30 and 70 which includes the early plateau phase before the
stream signal becomes important is $\dot P_+ = -1.8\times 10^{-4}\ \rm
s\ s^{-1}$.

Between cycles $\sim$110 and 150, the O-C phase diagrams in Figures
\ref{fig: sh1omc} and \ref{fig: sh2omc} show phase shifts of $\sim$0.5
cycles.  This is the result of the continued fading of the disk
superhump, and the transition to the stream/late superhump signal.
Careful inspection of the top panels of Figures \ref{fig: sh1omc} and
\ref{fig: sh2omc} near days 68 and 174 in fact shows the decreasing
amplitude of the disk superhump, and the relatively constant amplitude
of the stream superhump.   By cycle $\sim150$ (days  $\sim$72 and
176), the disk superhump amplitude is negligible, and all that remains
is the signal from the stream superhump.  The smoothed SAP light curve
shown in the top panel shows that these times correspond
to the return to the quiescent state during which the global viscosity
is again low.  It is also interesting that $\dot P_+$ itself appears
to be increasing relatively linearly during much of the plateau phase
with an average rate of $\ddot P \sim$$10^{-9}\rm\ s^{-1}$.  At present this is
not explained by the numerical simulations.  It may simply be that
this result reflects the growing relative importance of the stream
superhump signal on the phase of the 3-cycle sine fit.  This is almost
certainly the case during the period peaks found at days $\sim$71 and
175, where we find that the sine fits are pulled to longer period by
the complex and rapidly changing waveform (e.g., Figure \ref{fig: trans}).

In the quiescent interval before the first subsequent outburst the O-C
diagram shows a concave-downward shape indicating a negative
$\pdotplus \sim -2\times10^{-4}\ \rm s\ s^{-1}$.  We speculate that
the behavior of the O-C curve in response to the outburst following the
first superoutburst may indicate that the outburst may effectively expand
the radius of the disk causing a faster apsidal precession.
Unfortunately, there is a gap in the \Kepler data that starts just
after the initial rise of the outburst following the second superoutburst.
The value of $\pdotplus$ averaged over the the last 2 measured bins
for both superoutbursts is $\pdotplus \sim -3\times10^{-4}\ \rm s\
s^{-1}$.

The measured values of $\pdotplus$ for V344 Lyr are consistent with
those reported in the extensive compilation of \citet{kato09}. To make
a direct comparison with Kato et al., who calculate $\pdotplus$ over
the first 200 cycles (i.e., plateau phase), we average all the
$\pdotplus$ measurements out to the drop to quiescence, and find an
average value of $-6\times10^{-5} \ \rm s\ s^{-1}$ for the first
superoutburst and $-9\times10^{-5} \ \rm s\ s^{-1}$ for the second.
These values for V344 Lyr are entirely consistent with the Kato et
al.\ results as shown in their Figure 8.

In \citet{still10} we noted that V344 Lyr was unusual (but not unique)
in that superhumps persist into quiescence and through the following
outburst in Q2.  Other systems that have been observed to show
(late) superhumps into quiescence more typically have short orbital periods,
including V1159 Ori \citep{patterson95}, ER UMa \citep{gao99,zhao06},
WZ Sge \citep{patterson02wzsge}, and the WZ Sge-like star V466 And
\citep{chochol10}, among others.  The identification of late
superhumps is a matter of contention in some cases \citep{kato09},
and the post-superoutburst coverage of targets is more sparse than the
coverage during superoutbursts.  Thus it is difficult to know if
post-superoutburst superhumps are common or rare at this time.

\subsection{Observed Negative Superhumps}

\begin{figure}
\epsscale{1.0}
\plotone{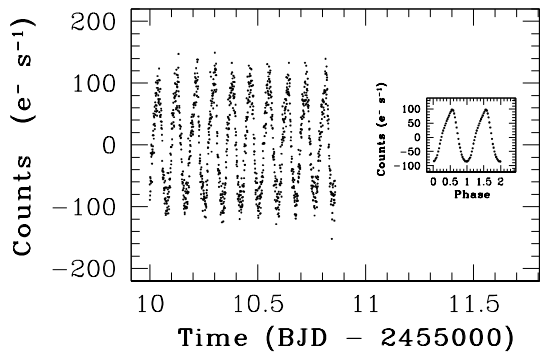}
\caption{The Figure shows 10 cycles of the negative superhump signal.
The inset shows the average pulse shape over days 5 to 25.  The signal
is sawtoothed with a rise time roughly twice the fall time, and
appears consistent with the ray-traced SPH simulation pulse shapes
published by Wood \& Burke (2007).}
\label{fig: aveneglc} 
\end{figure}

As noted above in \S 2.2, the 2.06-hr (11.4 c/d) signal that dominates
the light curve for the first  $\sim$35 days of Q2 is now understood
to be the result of a negative superhump.  This yields a value for the
period deficit (Equation \ref{eq: eps-}) of $\epsilon_- = 2.5$\%.  The
maximum amplitude at quiescence is $A\sim0.8$ mag. 
Figure~\ref{fig: aveneglc} shows 10 cycles of the negative superhump
signal during this time.  The inset shows the mean pulse shape
averaged over days 5 to 25 (roughly 230 cycles).  The signal is
approximately sawtoothed with a rise time roughly twice the fall time.
It appears consistent with the pulse shapes \citet{wb07} obtained
using ray-trace techniques on 3D simulations of tilted disks (their
Figure 3).  Negative superhumps dominate the power in days $\sim$2--35
and again in days $\sim$100--160.  

The signal observed near the beginning of Q2 reveals a remarkably
large rate of period change -- large enough that it can be seen in the
harmonics of the Fourier transform shown in Figure~\ref{fig: 2dDFT} as
a negative slope towards lower frequency with time.  A nonlinear least
squares fit to the fundamental period measured during days 2.5-7.5
yields $P_-=2.05006\pm0.00005$ hr.  A fit to the data from days
22--26, however, yields $P_-=2.06273\pm0.00005$ hr. The formal errors
from non-linear least squares fits underestimate the true errors by as
much an order of magnitude \citep{mo99}, but even if this is the case,
these two results differ by $\sim$25$\sigma$.  Taken at face value,
they yield a rate of period change of $\dot P_- \sim 3\times10^{-5}\rm\
s\ s^{-1}$.  Similarly, we fit the negative superhump periods in two
4-day windows centered on days 112.0 and 121.0.  The periods obtained
from non-linear least squares are $P_- = 2.0530 \pm 0.0002$ hr and $P_- =
2.066038 \pm 0.00008$ hr, respectively, which yields
$\dot P_- \sim 6\times10^{-5}\rm\ s\ s^{-1}$ over this time span.
In their recent comprehensive analysis of the evolution of CVs as
revealed by their donor stars, \citet{knigge11} estimate that for
systems with $\Porb\sim2$ hr the rate of orbital period change should
be $\dot \Porb\sim-7\times10^{14}\rm\ s\ s^{-1}$ (see their Figure
11).  Clearly the $\sim$2.06-hr signal cannot be orbital in origin.
In some negatively superhumping systems with high inclinations, the
precessing tilted disk can modulate the mean brightness
\citep[e.g.][]{stanishev02}.  We found no significant signal in the
Fourier transform at the precession period of $\sim$3.6 d.

\begin{figure}
\epsscale{1.0}
\plotone{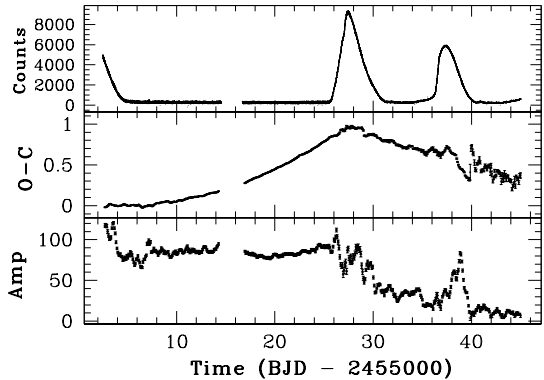}
\caption{The SAP light curve (top panel) O-C phase diagram (middle
panel) and amplitude (bottom panel) of the negative superhump signal
during days 2.5 to 45.0.  A 5-cycle sinusoid was fit to the data,
moving the window 1 cycle between fits.  The amplitude of the signal
is large and relatively constant initially, and the phase variations
are concave upward, indicating a lengthening period.   During outburst
1, there is a cusp in the O-C diagram, indicating a jump to
shorter period, and after outburst 1, the amplitude begins to decline.   }
\label{fig: negshomc} 
\end{figure}

In Figure~\ref{fig: negshomc} we show the results of the O-C analysis
for the Q2 data.  To create the Figure, we fit 5-cycle sine curves of
period 2.05 hr to the residual light curve, shifting the data by one
cycle between fits.  The shape of the O-C diagram is concave up until
the peak of the first outburst at day $\sim$28 indicating that the
period of the signal is lengthening during this time span.  The
magnitude of the negative superhump period deficit is inversely
related to the retrograde precession period of the tilted disk -- a
shorter precession period yields a larger period deficit.  A disk that
was not precessing at all would show a negative superhump period equal
to the orbital period.  The observation that the negative superhump
period in V344 Lyr is lengthening during days $\sim$2 to 27 indicates
that the precession period of the tilted disk is increasing (i.e., the rate
of precession is decreasing).
Coincident with the first DN outburst (outburst 1) in Q2, there is a cusp
in the O-C diagram, indicating a jump to shorter period (faster
retrograde precession rate).  The amplitude of the signal begins to
decline significantly following outburst 1, and the signal is effectively
quenched by outburst 2.  Note that between days $\sim$28 and 35 the O-C
diagram is again concave up, although with less curvature than before
outburst 1.

\begin{figure*}
\plotone{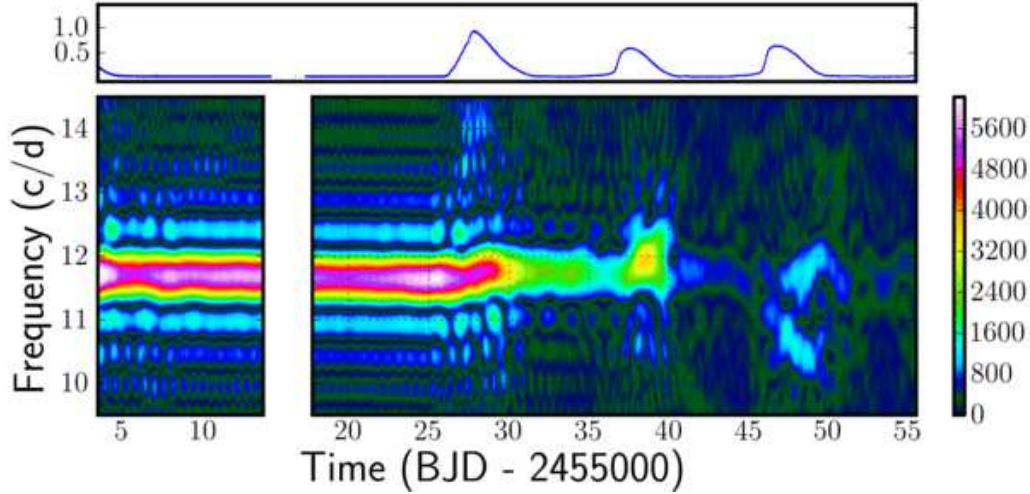}
\caption{The 2D DFT of the pre-superoutburst Q2 data.  Shown are 
amplitude per cadence versus time and frequency.  The window
width was 2 days, and the window was shifted  1/8 day between 
transforms. Most obvious is that the first outburst shifts the power
to higher frequency, but careful inspection also shows that the
frequency evolves towards longer period up to day $\sim$25 (e.g., note
the slope of the sidelobe at $\sim$11 c/d with respect to the grid
line).  Outbursts 2 and 3 both appear to generate power on the decline
from maximum.}
\label{fig: 2dDFTq2}
\end{figure*}

We show the 2D DFT of the pre-superoutburst Q2 data in
Figure~\ref{fig: 2dDFTq2}.   Here we used a window width of 2 days
that was shifted 1/8 day between transforms. We plot the amplitude in
counts per cadence.  It is evident that outburst 1 shifts the oscillation
frequency, as well quenching the amplitude of the signal.  Outburst 2
triggers a short-lived signal with a period of roughly 11.9 c/d (2.02
hr), and outburst 3 appears to generate signals near the frequencies of the
negative and positive superhumps that rapidly evolve to higher and
lower frequencies, respectively, only to fade to to noise background
by the end of the outburst.  Outburst 3 has a somewhat slower rise to
maximum than most of the outbursts in the time series and is the last
outburst
before the first superoutburst, but is otherwise unremarkable.  This
is the only time we see this behavior in the 3 quarters of data we
present, so it is unclear what the underlying physical mechanism is.

\begin{figure*}
\plotone{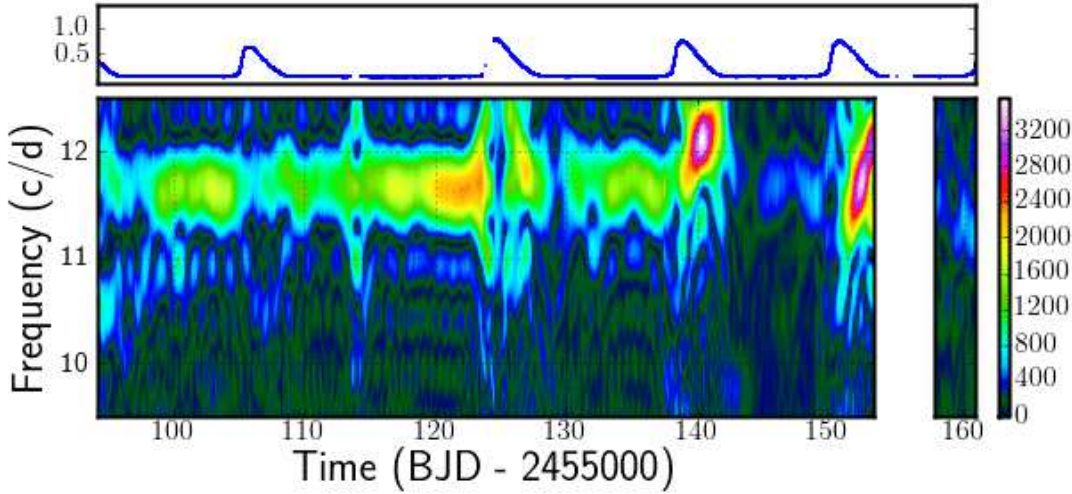}
\caption{The 2D DFT of the inter-superoutburst Q3 data.  Shown are 
amplitude per cadence versus time and frequency.  The window
width was 2 days, and the window was shifted  1/8 day between 
transforms. The negative superhump signal is present until day
$\sim$140.  Higher-frequency power is generated on the decline from
maximum of the last 2 outbursts.}
\label{fig: 2dDFTq3}
\end{figure*}

Although much of the Q3 light curve is dominated by the negative
superhump signal, the amplitude is much lower than early in Q2, and in
addition there is contamination from the orbital and positive
superhump signals.  In
Figure~\ref{fig: 2dDFTq3} we show the 2D DFT for the Q3 data between
days 93 and 162, again showing the amplitude in counts per cadence
versus time and frequency.  We used a window width of 2 days that was
shifted 1/8 day between transforms. 

\begin{figure}
\epsscale{1.0}
\plotone{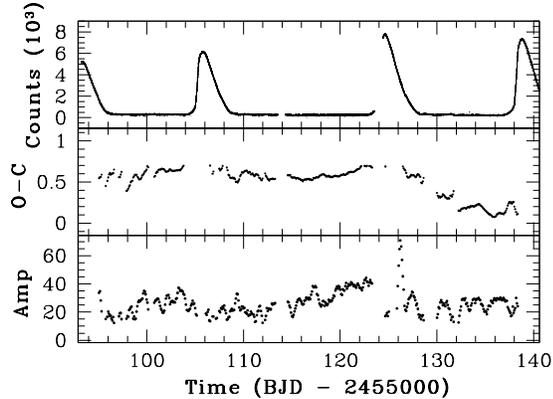}
\caption{The SAP light curve (top panel) O-C phase diagram (middle
panel) and amplitude (bottom panel) of the negative superhump signal
during days 106.5 to 123.2.  As in Figure \ref{fig: negshomc}, the shape of the O-C curve is concave upward, indicating a positive rate of period change. }
\label{fig: negshomc2} 
\end{figure}

In Figure \ref{fig: negshomc2} we show the O-C phase diagram obtained
by fitting a 5-cycle sine curve of period 2.06 hr to data spanning
days 93.2 to 140.0.  The amplitude during this time is considerably
smaller than was the case for the Q2 negative superhumps.
Before day 106, there appears to be contamination from periodicities
near the superhump frequency of 10.9 c/d that are evident in 
Figure~\ref{fig: 2dDFTq3}, and after day 126 the signal fades
dramatically. 
It was only during days 106.5 to 123.2 that the
amplitude of the negative superhump signal was large enough, stable
enough, and uncontaminated to yield a clean O-C phase diagram.  These
data lie between outbursts 8 and 9, and comprise the longest
quiescent stretch in Q3.  It can be seen that the O-C curve is again
concave upward indicating a positive rate of period change as
calculated above, and the bottom panel indicates that the amplitude of
the signal is increasing during this time span.

The retrograde precession rate of a tilted accretion disk is a direct
function of the effective (mass weighted) radius of the disk.  
Several groups have studied the precession properties of tilted disks
\citep{papterq95,larwood96,larwood98,lp97,lai99}.  
\citet{papaloizou97} derived the following expression for the induced
precession frequency $\omega_p$ of a tilted accretion disk,
\begin{equation}
\omega_p = -{3\over 4}{GM_2\over a^3}
  {\int\Sigma r^3\, dr\over \int \Sigma\Omega r^3\, dr}\,\cos\delta
\label{eq: pt95}
\end{equation}
where 
 $\omega_p$ is the leading-order term of the induced precession
frequency for a differentially rotating fluid disk, calculated using
linear perturbation theory,
$\Sigma(r)$ is the axisymmetric surface density profile 
and $\Omega(r)$ the unperturbed Keplerian angular velocity profile,
$a$ is the orbital separation, $M_2$ is the mass of the secondary, 
and $\delta$ is the tilt of the disk
with respect to the orbital plane.  The integrals are to be taken
between the inner and outer radii of the disk.

In a later study of the precession of tilted accretion disks, 
\citet[][and see Larwood (1998)]{larwood97} derived the expression for
the precession frequency of a disk with constant
surface density $\Sigma$ and polytropic equation of state with ratio
of specific heats equal to 5/3:
\begin{equation}
{\omega_p\over\Omega_0} = -{3\over 7}q\left({R_0\over
a}\right)^3\cos\delta,
\label{eq: larwood}
\end{equation}
where here $\Omega_0$ is the Keplerian angular
velocity of the outer disk of radius $R_0$, and $q$ is the mass ratio.

The physical interpretation of Equations \ref{eq: pt95} and \ref{eq:
larwood} is that tilted accretion disks weighted to larger radii will
have higher precession frequencies than those weighted to smaller
radii.  For example, if we have 2 disks with the same nominal tilt and
total mass, where one has a constant surface density and the other
with a surface density that increases with radius, the second disk
will have a higher precession rate, and would yield a negative
superhump frequency higher than the first.  A third disk with most of
its mass concentrated at small radius would have a lower
precession frequency and yield a negative superhump signal nearest the
orbital signal. 

In this picture the increasing precession period indicated by the
positive rate of period change for the negative superhump signal
$\dot P_-$ might at first seem counter-intuitive since the
disk is gaining mass at quiescence.  However, the key fact is that
tilted disks accrete most of their mass at {\it small} radii, since
the accretion stream impacts the face of the tilted disk along the
line of nodes \citep{wb07,wts09}.  The accretion stream impacts the rim of
the disk only twice per orbit (refer back to Figure~\ref{fig:
sph-}).  Thus, the effective (mass weighted)
radius of an accreting tilted disk {\it decreases} with time, causing
a slowing in the retrograde precession rate $\omega_p$, and an
increase in the period of the negative superhump signal $P_-$.

A detailed analysis of the data, theory, and numerical model results
should allow us to probe the time evolution of the mass distribution
in disks undergoing negative superhumps, and hence the low-state
viscosity mechanism.  The unprecedented quality and quantity of the
\Kepler time series data suggests that V344 Lyr and perhaps other
\Kepler-field CVs that display negative superhumps may significantly
advance our understanding of the evolution of the mass distribution in
tilted accretion disks.   

The cause of disk tilts in CVs is still not satisfactorily explained.
In the low-mass x-ray binaries it is believed that radiation pressure
can provide the force necessary to tilt the disk out of the orbital
plane  \citep{petterson77,ip90,foulkes06,ip08},
however this mechanism is not effective in the CV scenario.
\citet{bow88} suggested in their work on TV Col that magnetic fields
near the L1 region might deflect the accretion stream out of the
orbital plane, but as noted in \citet{wb07} the orbit-averaged angular
momentum vector of a deflected stream would still be parallel to the
orbital angular momentum variable.  \citet{murrayea02} demonstrated
numerically that a disk tilt could be generated by instantaneously
turning on a magnetic field on the secondary star.  Although their
tilt decayed with time (the orbit-averaged angular momentum argument
again), their results suggest that changing magnetic field geometries
could generate disk tilt.  Assuming that the disk viscosity is
controlled by the MRI \citep{bh98,balbus03}, it is plausible that
differentially-rotating plasmas may also be subject to magnetic
reconnection events (flares) which are asymmetrical with respect to the disk
plane, or that during an outburst the intensified disk field may
couple to the tilted dipole field on the primary star
\citep[e.g.,][]{lai99} or the field of the secondary star
\citep{murrayea02}.

\begin{figure}
\epsscale{1.0}
\plotone{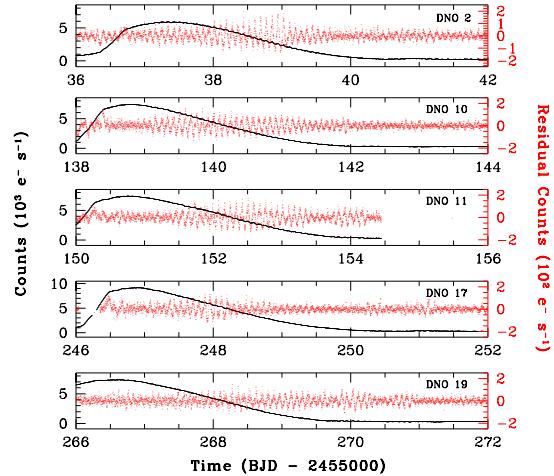} 
\caption{Dwarf nova outbursts 2, 10, 11, 17, and 19, each showing
oscillations on the decline from maximum.  The black points show the
SAP light curves, summed two points for clarity. The red points show
the residual light curves, also summed two points.}
\label{fig: dnofig} 
\end{figure}

With these ideas in mind, the behavior of V344 Lyr during outbursts 
2, 10, 11, 17, and 19 is tantalizing.  First, again consider the 2D
DFTs from Q2, Q3, and Q4 shown in Figures~\ref{fig: 2dDFTq2},
\ref{fig: 2dDFTq3} and \ref{fig: 2dDFTq4}, respectively.  In each of
these cases, there is power generated at a frequency consistent with
the negative superhump frequency on the decline from maximum light.
Outbursts 2 and 10 appear to excite a frequency of roughly 12 c/d
($\sim$2 hr), outburst 17 excites the negative superhump frequency for
$\sim$3 days, and outbursts 11 and 19 appear to excite power at the
negative superhump frequency that rapidly evolves to shorter
frequencies.  We show the SAP light curves for these outbursts as well
as the residual light curves in Figure~\ref{fig: dnofig}.  The
residual light curves for these 5 outbursts all appear to show the
excitation of a frequency near or slightly greater than the negative 
superhump frequency that dominates early in Q2.  This is about 1/3 of
the normal outbursts in the 3 quarters of \Kepler data -- the other 12
outbursts
do not show evidence for having excited new frequencies.   Thus, while
additional data are clearly required and our conclusions are
speculative, we suggest that these results
support a model in which the disk tilt is generated by
the transitory (impulsive) coupling between an
intensified disk magnetic field and the field of the primary or
secondary star.  The fact that these 5 outburst events yield frequencies
near 12 c/d appears to support the model that it is the mass in the 
outer disk that is initially tilted out of the plane.

\section{Conclusions}

We present the results of the analysis of 3 quarters of \Kepler time
series photometric data from the system V344 Lyr.  Our major findings
are:

\begin{enumerate}

\item The orbital, negative superhump, and positive superhump periods
are $\Porb=2.11$ hr, $P_- = 2.06$ hr, and $P_+ = 2.20$ hr, giving a
positive superhump period excess of $\epsilon_+ = 4.4$\%, and a
negative superhump period deficit of $\epsilon_- = 2.5$\%.

\item The quality of the \Kepler data is such that we can constrain
significantly the models for accretion disk dynamics that have been
proposed in the past several years.

\item The evolution of the pulse shapes and phases of the positive
superhump residual light curve provides convincing evidence in support
of the two-source model for positive superhumps.  Early in the
superoutburst, viscous dissipation in the strongly convergent flows of
the flexing disk provide the modulation observed at the superhump
frequency.  Once the system has returned to quiescence, the modulation
is caused by the periodically-variable dissipation at the bright spot
as it sweeps around the rim of the still non-axisymmetric, flexing
disk.  During the transition the O-C phase diagram shows a shift of
$\sim0.5$ in phase.

\item Superoutbursts begin as normal DN outbursts.  The rise to
superoutburst is largely explained by the thermal-viscous limit cycle
model discussed in Paper II.  Beyond this
luminosity source which does a reasonable job of matching the lower
envelope of the superoutburst light curve, there is additional
periodic dissipation that generates the superhump signals.  The
sources of the periodic dissipation are (i) the strongly convergent
flows that are generated once per superhump cycle as the disk is
compressed in the radial direction opposite the secondary, and (ii)
the variable depth of the bright spot as it sweeps around the rim of
the non-axisymmetric oscillating disk.

\item Numerical experiments that individually isolate the two proposed
physical sources of the positive superhump signal yield results that
are broadly consistent with the signals in the data.

\item The positive superhumps show significant changes in period that
occur in both superoutbursts.  The average $\dot P_+ \sim
6\times10^{-5}\rm\ s\ s^{-1}$ for the first superoutburst and  $\dot
P_+ \sim 9\times10^{-5}\rm\ s\ s^{-1}$ for the second are consistent
with literature results.  The data reveal that $\dot P_+$ itself
appears to be increasing relatively linearly during much of the
plateau phase at an average rate for the two superoutbursts of $\ddot
P \sim$$10^{-9}\rm\ s^{-1}$.  

\item The negative superhumps show significant changes in period with
time, resulting from the changing mass distribution (moment of
inertia) of the tilted disk.  As the mass of the inner disk increases before
outburst 1, the retrograde precession period increases, consistent with
theoretical predictions.  These data are rich with unmined
information.

\item Negative superhumps appear to be excited as a direct result of
some of the dwarf nova outbursts.  We speculate that the
MRI-intensified disk field can couple to the field of the primary or
secondary star and provide an impulse that tilts the disk out of the
orbital plane.  Continued monitoring by \Kepler promises to shed light
on this important unsolved problem.

\end{enumerate}

The system V344 Lyr continues to be monitored at short cadence by the
\Kepler mission.  It will undoubtedly become the touchstone system
against which observations of all other SU UMa CVs will be compared,
as the quantity and quality of the time series data are unprecedented
in the history of the study of cataclysmic variables.  The \Kepler
data for V344 Lyr promise to reveal details of the micro- and macrophysics 
of stellar accretion disks that would be impossible
 to obtain from ground-based observations.

\acknowledgments

\Kepler was selected as the 10th mission of the Discovery Program.
Funding for this mission is provided by NASA, Science Mission
Directorate.  All of the data presented in this paper were obtained
from the Multimission Archive at the Space Telescope Science Institute
(MAST).  STScI is operated by the Association of Universities for
Research in Astronomy, Inc., under NASA contract NAS5-26555. Support
for MAST for non-HST data is provided by the NASA Office of Space
Science via grant NAG5-7584 and by other grants and contracts. This
research was supported in part by the American Astronomical Society's
Small Research Grant Program in the form of page charges.  We
thank Marcus Hohlmann from the Florida Institute of Technology and the
Domestic Nuclear Detection Office in the Dept. of Homeland Security
for making computing resources on a Linux cluster available for this
work. We thank Joseph Patterson of Columbia University for sending us
the data used in Figure 19 in electronic form.

{\it Facilities:} \facility{\Kepler}


\begin{thebibliography}{}

\bibitem[Ak et al.(2008)]{ak08} Ak, T., Bilir, S., Ak, S., \& Eker, Z.\ 2008, New Astronomy, 13, 133 

\bibitem[Balbus(2003)]{balbus03} Balbus, S.~A.\ 2003, \araa, 41, 555

\bibitem[Balbus \& Hawley(1998)]{bh98} Balbus, S.~A., \& Hawley, J.~F.\ 1998, 
	Reviews of Modern Physics, 70, 1 
	
\bibitem[Barrett et al.(1988)]{bow88} Barrett, P., 
	O'Donoghue, D., \& Warner, B.\ 1988, \mnras, 233, 759 

\bibitem[Bildsten et al.(2006)]{bild06} Bildsten L., Townsley D.~M., Deloye C.~J.,
	Nelemans G., 2006, ApJ, 640, 466

\bibitem[Bonnet-Bidaud et al.(1985)]{bbmm85} Bonnet-Bidaud, J.~M.,
	Motch, C., \& Mouchet, M.\ 1985, \aap, 143, 313 



\bibitem[Borucki et al.(2010)]{borucki10} Borucki, W.~J., et al.\
	2010, Science, 327, 977 

\bibitem[Caldwell et al.(2010)]{caldwell10} Caldwell, D.~A., et al.\
	2010, \apjl, 713, L92 

\bibitem[Cannizzo(1993a)]{cannizzo93} Cannizzo, J. K. 1993, Accretion
	Disks in Compact Stellar Systems, ed. J. C. Wheeler (Singapore: World
	Scientific), 6

\bibitem[Cannizzo(1993b)]{cannizzo93ss} Cannizzo, J. K. 1993, ApJ, 419, 318

\bibitem[Cannizzo(1998)]{cannizzo98} Cannizzo, J. K. 1998, ApJ, 494, 366

\bibitem[Cannizzo et al.(2010)]{cannizzo10} 
	Cannizzo, J.~K., Still, M.~D., Howell, S.~B., Wood, M.~A., 
 	\& Smale, A.~P.  
      	2010, ApJ, 725, 1393

\bibitem[Cannizzo et al.(2011)]{cannizzo11} 
	Cannizzo, J.~K., Smale, A. P., Still, M.~D., Wood, M.~A., 
	\& Howell, S. B.
	2011, ApJ, submitted
	
\bibitem[Charles et al.(1991)]{charlesea91} Charles, P.~A.,
	Kidger, M.~R., Pavlenko, E.~P., Prokof'eva, V.~V., \& Callanan, P.~J.\
	1991, \mnras, 249, 567 

\bibitem[Chochol et al.(2010)]{chochol10} Chochol, D., Katysheva, 
	N.~A., Shugarov, S.~Y., Volkov, I.~M., 
	\& Andreev, M.~V.\ 2010, Contributions of the Astronomical Observatory
	Skalnate Pleso, 40, 19 

\bibitem[Faulkner et al.(1972)]{ffw72} Faulkner, J., 
	Flannery, B.~P., \& Warner, B.\ 1972, \apjl, 175, L79

\bibitem[Feldmeier et al.(2011)]{feldmeier11} Feldmeier, J.~J., et 
	al.\ 2011, \apj, in press (arXiv:1103.3660)

\bibitem[Fontaine et al.(2011)]{fontaine11} Fontaine, G., et al.\ 
	2011, \apj, 726, 92 

\bibitem[Foulkes et al.(2006)]{foulkes06} Foulkes, S.~B., Haswell,
	C.~A., \& Murray, J.~R.\ 2006, \mnras, 366, 1399 

\bibitem[Frank et al.(2002)]{fkr02} Frank, J., King, A., 
	\& Raine, D.~J.\ 2002, Accretion Power in Astrophysics, by Juhan Frank
	and Andrew King and Derek Raine, pp.~398.~ISBN 0521620538.~Cambridge,
	UK: Cambridge University Press, February 2002.,  

\bibitem[Gao et al.(1999)]{gao99} Gao, W., Li, Z., Wu, X., 
	Zhang, Z., \& Li, Y.\ 1999, \apjl, 527, L55
	
\bibitem[Gilliland et al.(2010)]{gilliland10} Gilliland, R.~L., et 
	al.\ 2010, \pasp, 122, 131 

\bibitem[Haas et al.(2010)]{haas10} Haas, M. R., et al.\ 2010,
	ApJL, 713, L115
	
\bibitem[Harvey et al.(1998)]{harvey98} Harvey, D.~A., Skillman, 
	D.~R., Kemp, J., Patterson, J., Vanmunster, T., Fried, R.~E., 
	\& Retter, A.\ 1998, \apjl, 493, L105 

\bibitem[Hellier(2001)]{hellier01} Hellier, C. 2001,
    Cataclysmic Variable Stars: How and Why They Vary,
	Springer-Praxis Books in Astronomy \& Space Sciences: Praxis
	Publishing

\bibitem[Hessman et al.(1992)]{hessman92} Hessman, F.~V., Mantel, K.-H., Barwig, H., \& Schoembs, R.\ 1992, \aap, 263, 147 

\bibitem[Howell et al.(1996)]{howell96} Howell, S.~B., Reyes, 
	A.~L., Ashley, R., Harrop-Allin, M.~K., 
	\& Warner, B.\ 1996, \mnras, 282, 623 

\bibitem[Hynes et al.(2006)]{hynesea06} Hynes, R.~I., et al.\ 2006,
	\apj, 651, 401 

\bibitem[Iping \& Petterson(1990)]{ip90} 
	Iping R.~C., Petterson J.~A., 1990, A\&A, 239, 221

\bibitem[Ivanov \& Papaloizou(2008)]{ip08} 
	Ivanov P.~B., Papaloizou J.~C.~B., 2008, MNRAS, 384, 123


\bibitem[Jenkins et al.(2010)]{jenkins10} Jenkins, J.~M., et al.\ 
	2010, \apjl, 713, L87 


\bibitem[Kato(1993)]{kato93} Kato, T.\ 1993, \pasj, 45,
	L67

\bibitem[Kato et al.(2002)]{kato02} Kato, T., Poyner, G.,
	\& Kinnunen, T.\ 2002, \mnras, 330, 53 
	
\bibitem[Kato et al.(2009)]{kato09} Kato, T., et al.\ 2009, 
\pasj, 61, 395 


\bibitem[Kato et al.(2010)]{kato10} Kato, T., et al.\ 2010, 
\pasj, 62, 1525 

\bibitem[Kim et al.(2009)]{kim09} Kim, Y., Andronov, I.~L., Cha, S.~M.,
	Chinarova, L.~L., \& Yoon, J.~N.\ 2009, \aap, 496, 765

\bibitem[Knigge et al.(2011)]{knigge11} Knigge, C., Baraffe, I., \&
	Patterson, J.\ 2011, \apjs, 194, 28 

\bibitem[Koch et al.(2010)]{koch10} Koch, D.~G., et al.\ 2010, 
	\apjl, 713, L79 
	
\bibitem[Kunze(2002)]{kunze02} Kunze, S.\ 2002, in ASP Conf. Series 261: The Physics of 
	Cataclysmic Variables and Related Objects, eds. B.T. G\"ansicke, K. Beuermann, \& K. Reinsch, 497 


\bibitem[Kunze(2004)]{kunze04} Kunze, S.\ 2004, Revista 
	Mexicana de Astronomia y Astrofisica Conference Series, 20, 130 

\bibitem[Lai(1999)]{lai99} Lai, D.\ 1999, \apj, 524, 1030

\bibitem[Larwood(1997)]{larwood97} Larwood, J.~D.\ 1997, \mnras, 290, 490 

\bibitem[Larwood(1998)]{larwood98} Larwood, J.\ 1998, \mnras, 299, L32 

\bibitem[Larwood et al.(1996)]{larwood96} Larwood, J.~D., Nelson, 
	R.~P., Papaloizou, J.~C.~B., \& Terquem, C.\ 1996, \mnras, 282, 597 

\bibitem[Larwood 
\& Papaloizou(1997)]{lp97} Larwood, J.~D., \& Papaloizou, J.~C.~B.\ 1997, \mnras, 285, 288 

\bibitem[Lasota(2001)]{lasota01} Lasota, J.-P. 2001, New Astron. Rev.,
	45, 449

\bibitem[Mineshige et al.(1992)]{mho92} Mineshige, S., Hirose, M., \&
	Osaki, Y.\ 1992, \pasj, 44, L15 

\bibitem[Montgomery 
\& Odonoghue(1999)]{mo99} Montgomery, M.~H., \& Odonoghue, D.\ 1999, Delta Scuti Star Newsletter, 13, 28 

\bibitem[Murray et al.(2002)]{murrayea02} Murray J.~R., 
	Chakrabarty D., Wynn G.~A., Kramer L., 2002, MNRAS, 335, 247



\bibitem[Nelemans(2005)]{nelemans05} Nelemans, G.\ 2005, in ASP Conf.
	Ser. 330, The Astrophysics of Cataclysmic Variables and Related
	Objects, ed. J.-M. Hameury \& J.-P. Lasota (San Francisco: ASP), 27

\bibitem[Nelemans et al.(2001)]{nelemans01} Nelemans, G., Steeghs, 
	D., \& Groot, P.~J.\ 2001, \mnras, 326, 621 

\bibitem[O'Donoghue \& Charles(1996)]{oc96} O'Donoghue,
	D., \& Charles, P.~A.\ 1996, \mnras, 282, 191 

\bibitem[Osaki(1985)]{osaki85} Osaki, Y.\ 1985, \aap, 144, 369 
	
\bibitem[Osaki(1989)]{osaki89} Osaki, Y.\ 1989, \pasj, 41, 1005 
		
\bibitem[Papaloizou et al.(1997)]{papaloizou97} Papaloizou, 
J.~C.~B., Larwood, J.~D., Nelson, R.~P., 
\& Terquem, C.\ 1997, Accretion Disks - New Aspects, 487, 182 

\bibitem[Papaloizou \& Terquem(1995)]{papterq95} Papaloizou, J.~C.~B., \&
	Terquem, C.\ 1995, \mnras, 274, 987

\bibitem[Patterson(1999)]{patterson99} Patterson, J.\ 1999, in Disk 
	Instabilities in Close Binary Systems, eds. S. Mineshige and 
	J. C. Wheeler, (Kyoto: Universal Acad. Press), 61 

\bibitem[Patterson et al.(1993)]{pattersonea93a} Patterson, J., 
	Halpern, J., \& Shambrook, A.\ 1993, \apj, 419, 803 
	
\bibitem[Patterson et al.(1995)]{patterson95} Patterson, J., 
	Jablonski, F., Koen, C., O'Donoghue, D., 
	\& Skillman, D.~R.\ 1995, \pasp, 107, 1183 

\bibitem[Patterson et al.(2000)]{patterson00} Patterson, J., Kemp, 
	J., Jensen, L., Vanmunster, T., Skillman, D.~R., Martin, B., Fried, R., 
	\& Thorstensen, J.~R.\ 2000, \pasp, 112, 1567

\bibitem[Patterson et al.(1992)]{patterson92} Patterson, J., 
	Sterner, E., Halpern, J.~P., \& Raymond, J.~C.\ 1992, \apj, 384, 234 

\bibitem[Patterson et al.(1993)]{pattersonea93b} Patterson, J., 
	Thomas, G., Skillman, D.~R., \& Diaz, M.\ 1993, \apjs, 86, 235 

\bibitem[Patterson et al.(2002a)]{patterson02} Patterson, J., et 
	al.\ 2002, \pasp, 114, 65

\bibitem[Patterson et al.(2002b)]{patterson02wzsge} Patterson, J.,
	et al.\ 2002, \pasp, 114, 721 

\bibitem[Patterson et al.(2003)]{patterson03} Patterson, J.,
	et al.\ 2003, \pasp, 115, 1308 
	
\bibitem[Patterson et al.(2005)]{patterson05} Patterson, J.,
	et al.\ 2005, \pasp, 117, 1204 

\bibitem[Petterson(1977)]{petterson77}
	Petterson J.~A., 1977, ApJ, 216, 827

\bibitem[Provencal et al.(1995)]{provencal95} Provencal, J.~L., et 
	al.\ 1995, \apj, 445, 927 

\bibitem[Retter et al.(1997)]{retterea97} Retter, A.,
	Leibowitz, E.~M., \& Ofek, E.~O.\ 1997, \mnras, 286, 745 

\bibitem[Retter et al.(2002)]{retterea02} Retter, A., Chou, Y.,
	Bedding, T.~R., \& Naylor, T.\ 2002, \mnras, 330, L37 

\bibitem[Roelofs et al.(2007)]{roelofs07} Roelofs, G.~H.~A., Groot, P.~J., 
	Nelemans, G., Marsh, T.~R., \& Steeghs, D.\ 2007, \mnras, 379, 176
		
\bibitem[Rolfe et al.(2001)]{rolfe01} Rolfe, D.~J., Haswell, 
C.~A., \& Patterson, J.\ 2001, \mnras, 324, 529 

\bibitem[Schoembs(1986)]{schoembs86} Schoembs, R.\ 1986, \aap, 158, 233
	
\bibitem[Simpson \& Wood(1998)]{sw98} Simpson, J.~C., \& Wood, M.~A.\
	1998, \apj, 506, 360 

\bibitem[Skillman et al.(1997)]{skillmanea97} Skillman, D.~R., Harvey,
	D., Patterson, J., \& Vanmunster, T.\ 1997, \pasp, 109, 114 

\bibitem[Skillman et al.(1999)]{skillman99} Skillman, D.~R., 
Patterson, J., Kemp, J., Harvey, D.~A., Fried, R.~E., Retter, A., Lipkin, 
Y., \& Vanmunster, T.\ 1999, \pasp, 111, 1281

\bibitem[Smak(1967)]{smak67} Smak, J.\ 1967, Acta Astron., 17, 255 

\bibitem[Smak(2007)]{smak07} Smak, J.\ 2007, Acta Astron., 57, 87 

\bibitem[Smak(2008)]{smak08} Smak, J.\ 2008, Acta Astron., 58, 55 

\bibitem[Smak(2009)]{smak09} Smak, J.\ 2009, Acta Astron., 59, 121

\bibitem[Smak(2010)]{smak10} Smak, J.\ 2010, Acta Astron., 60, 357 
	
\bibitem[Smak(2011)]{smak11} Smak, J.\ 2011, Acta Astron., 61, 59

\bibitem[Smith et al.(2007)]{smith07} Smith, A.~J., Haswell, 
C.~A., Murray, J.~R., Truss, M.~R., 
\& Foulkes, S.~B.\ 2007, \mnras, 378, 785 

\bibitem[Solheim(2010)]{solheim10} Solheim, J.-E.\ 2010, \pasp, 
	122, 1133 

\bibitem[Stanishev et al.(2002)]{stanishev02} Stanishev, V.,
	Kraicheva, Z., Boffin, H.~M.~J., \& Genkov, V.\ 2002, \aap, 394, 625 

\bibitem[Sterken et al.(2007)]{sterken07} Sterken, C., Vogt, N., Schreiber,
	M.~R., Uemura, M., \& Tuvikene, T.\ 2007, \aap, 463, 1053

\bibitem[Still et al.(2010)]{still10} Still, M., Howell, S.~B., 
	Wood, M.~A., Cannizzo, J.~K., \& Smale, A.~P.\ 2010, 
	\apjl, 717, L113 

\bibitem[Templeton et al.(2006)]{templeton06} Templeton, M.~R., et 
	al.\ 2006, \pasp, 118, 236
		
\bibitem[Van Cleve(2010)]{vancleve10.6} Van Cleve, J., ed. 2010,
	Kepler Data Release Notes 6, KSCI-019046-001.

\bibitem[Vogt(1982)]{vogt82} Vogt, N.\ 1982, \apj, 252, 653

\bibitem[Warner(1995a)]{warner95} Warner, B. 1995a, Cataclysmic Variable
	Stars (Cambridge: Cambridge University Press)

\bibitem[Warner(1995b)]{warner95amcvn} Warner, B.\ 1995b, \apss,
	225, 249

\bibitem[Whitehurst(1988)]{whitehurst88} Whitehurst, R.\ 1988, 
\mnras, 232, 35 

\bibitem[Wood et al.(2005)]{wood05} Wood, M.~A., et al.\ 2005, 
	\apj, 634, 570  

\bibitem[Wood \& Burke(2007)]{wb07} Wood, M.~A., \& Burke, C.~J.\
	2007, \apj, 661, 1042 

\bibitem[Wood et al.(1986)]{wood86} Wood, J., Horne, K., Berriman, G.,
	Wade, R., O'Donoghue, D., \& Warner, B.\ 1986, \mnras, 219, 629 

\bibitem[Wood et al.(2000)]{wms00} Wood, M.~A., Montgomery, M.~M., \&
	Simpson, J.~C.\ 2000, \apjl, 535, L39 


\bibitem[Wood et al.(2009)]{wts09} Wood, M.~A., Thomas, D.~M., \&
	Simpson, J.~C.\ 2009, \mnras, 398, 2110 

\bibitem[Woudt et al.(2009)]{woudt09} Woudt, P.~A., Warner, B., 
	Osborne, J., \& Page, K.\ 2009, \mnras, 395, 2177
	
\bibitem[Zhao et al.(2006)]{zhao06} Zhao, Y., Li, Z., Wu, X., 
	Peng, Q., Zhang, Z., \& Li, Z.\ 2006, \pasj, 58, 367

\end{thebibliography}
\end{document}